\documentclass{aa}  
\usepackage{graphicx}
\usepackage{txfonts}
\begin{document} 
   \titlerunning{Dust properties of Lyman break galaxies at $z\sim3$}
   \authorrunning{\'Alvarez-M\'arquez, Burgarella, Heinis et al.}

   \title{Dust properties of Lyman break galaxies at $z\sim3$}

   \subtitle{}

\author{J.~\'Alvarez-M\'arquez\inst{1}
\and D.~Burgarella\inst{1}
\and S.~Heinis\inst{2}
\and V.~Buat\inst{1}
\and B.~Lo~Faro\inst{1}
\and M.~B\'ethermin\inst{3}
\and C.~E.~L\'opez-Fort\'in\inst{1}
\and A.~Cooray\inst{4}
\and D.~Farrah\inst{5}
\and P.~Hurley\inst{6}
\and E.~Ibar\inst{7}
\and O.~Ilbert\inst{1}
\and A.M.~Koekemoer\inst{8}
\and B.C.~Lemaux\inst{1}
\and I.~P\'erez-Fournon\inst{9,10}
\and G.~Rodighiero\inst{11}
\and M.~Salvato\inst{12}
\and D.~Scott\inst{13}
\and Y.~Taniguchi\inst{14}
\and J.~D.~Vieira\inst{15,16}
\and L.~Wang\inst{17,18}
}

\institute{
$^{1}$
Aix-Marseille Universit\'e, CNRS, LAM (Laboratoire d'Astrophysique de Marseille) UMR7326, 13388, France\\
$^{2}$
University of Maryland, Dept. of Astronomy, College park, MD\\
$^{3}$
European Southern Observatory, Karl-Schwarzschild Str. 2, D85748 Garching, Germany\\
$^{4}$
Department of Physics and Astronomy, University of California, Irvine, CA 92697, USA\\
$^{5}$
Department of Physics, Virginia Tech, Blacksburg, VA 24061, USA\\
$^{6}$
Astronomy Centre, Department of Physics and Astronomy, University of Sussex, Falmer, Brighton BN1 9QH, UK\\
$^{7}$
Instituto de F\'isica y Astronom\'ia, Universidad de Valpara\'iso, Avda. Gran Breta\~{n}a 1111, Valpara\'iso, Chile\\
$^{8}$
Space Telescope Science Institute, 3700 San Martin Drive, Baltimore, MD 21218, USA\\
$^{9}$
Instituto de Astrof\'isica de Canarias (IAC), C/V\'ia L\'actea, s/n, E-38200, La Laguna, Tenerife, Spain\\
$^{10}$
Departamento  de  Astrof\'isica,  Universidad  de  La  Laguna, E-38206, La Laguna, Tenerife, Spain\\
$^{11}$
Dipartimento di Fisica e Astronomia, Universit\'a di Padova, vicolo dell’Osservatorio 3, I–35122 Padova, Italy\\
$^{12}$
Max-Planck-Institute f\"ur Plasma Physics, Boltzmann Strasse 2, Garching 85748, Germany\\
$^{13}$
Department   of   Physics   and   Astronomy,   University   of   British Columbia, Vancouver, BC V6T 1Z1, Canads\\
$^{14}$
Research Center for Space and Cosmic Evolution, Ehime University, Bunkyo-cho, Matsuyama 790-8577, Japan\\
$^{15}$
Department  of  Physics,  University  of  Illinois  UrbanaChampaign, 1110 W. Green Street, Urbana, IL6180124\\
$^{16}$
Astronomy Department, University of Illinois at UrbanaChampaign, 1002 W. Green Street, Urbana, IL61801\\
$^{17}$
SRON Netherlands Institute for Space Research, Landleven 12, 9747 AD, Groningen, The Netherlands\\
$^{18}$
Institute for Computational Cosmology, Department of Physics, University of Durham, South Road, Durham, DH1 3LE, UK.\\
\email{javier.alvarez@lam.fr}}

   \date{Received ; accepted }

% \abstract{}{}{}{}{} 
% 5 {} token are mandatory
 
  \abstract
  % context heading (optional)
{Since the mid-1990s, the sample of Lyman break galaxies (LBGs) has been growing thanks to the increasing sensitivities in the optical and in near-infrared telescopes for objects at $z > 2.5$. However, the dust properties of the LBGs are poorly known because the samples are small and/or biased against far-infrared or sub-mm observations.}
  % aims heading (mandatory)
{This work explores from a statistical point of view the far-infrared (far-IR) and sub-millimeter (sub-mm) properties of a large sample of LBGs at $z\sim3$ that cannot be individually detected from current far-infrared observations.}
  % methods heading (mandatory)
{We select a sample of 22,~000 LBGs at $2.5 < z < 3.5$ in the COSMOS field using the dropout technique. The large number of galaxies included in the sample allows us to split it in several bins as a function of UV luminosity ($L_{\mathrm{FUV}}$), UV continuum slope ($\beta_{\mathrm{UV}}$) and stellar mass ($M_{*}$) to better sample their variety. We stack in PACS (100 and 160~$\mu$m) images from PACS Evolution Probe survey (PEP), SPIRE (250, 350 and 500$\mu$m) images from the Herschel Multi-tied Extragalactic Survey (HerMES) programs and AzTEC (1.1 mm) images from the Atacama Submillimetre Telescope Experiment (ASTE). Our stacking procedure corrects the biases induced by galaxy clustering and incompleteness of our input catalogue in dense regions.}
  % results heading (mandatory)
{We obtain the full infrared spectral energy distributions (SED) of subsamples of LBGs and derive the mean IR luminosity as a function of $L_{\mathrm{FUV}}$, $\beta_{\mathrm{UV}}$ and $M_{*}$. The average IRX (or dust attenuation) is roughly constant over the $L_{\mathrm{FUV}}$ range, with a mean of 7.9 (1.8 mag). However, it is correlated with $\beta_{\mathrm{UV}}$, $A_{FUV} = (3.15\pm0.12) + (1.47\pm0.14) ~\beta_{\mathrm{UV}}$,  and stellar mass, $\log \left( IRX \right) =(0.84\pm0.11) \log \left( M_{*}/10^{10.35}\right) + 1.17\pm0.05$. We investigate using a statistically-controlled stacking analysis as a function of ($M_{*}$, $\beta_{\mathrm{UV}}$) the dispersion of the IRX-$\beta_{\mathrm{UV}}$ and IRX-$M_{*}$ plane. On the one hand, the dust attenuation shows a departure by up to 2.8 mag above the mean IRX-$\beta_{\mathrm{UV}}$ relation, when $\log(M_{*}[M_{\sun}])$ increases from 9.75 to 11.5 in the same $\beta_{\mathrm{UV}}$ bin. That strongly suggests that $M_{*}$ plays an important role in shaping the IRX-$\beta_{\mathrm{UV}}$ plane. On the other hand, the IRX-$M_{*}$ plane is less dispersed for variation in the $\beta_{\mathrm{UV}}$. However, the dust attenuation shows a departure by up to 1.3 mag above the mean IRX-$M_{*}$ relation, when $\beta_{\mathrm{UV}}$ increases from -1.7  to 0.5 in the same $M_{*}$ bin. The low stellar mass LBGs ($\log(M_{*}[M_{\sun}]) < 10.5$) and red $\beta_{\mathrm{UV}}$ ($\beta_{\mathrm{UV}} > -0.7$), 15\% of the total sample, present a large dust attenuation than the mean IRX-$M_{*}$, but they still are in agreement with the mean IRX-$\beta_{\mathrm{UV}}$ relation. We suggest that we have to combine both, IRX-$\beta_{\mathrm{UV}}$ and IRX-$M_{*}$, relations to obtain the best estimation of the dust attenuation from the UV and NIR properties of the galaxies ($L_{\mathrm{FUV}}$, $\beta_{\mathrm{UV}}$, $M_{*}$). Our results enable us to study the average relation between star-formation rate (SFR) and stellar mass, and we show that our LBG sample lies on the main sequence of star formation at $z\sim3$. we demonstrate that the SFR is underestimate for LBGs with high stellar mass, but it give a good estimation for LBGs with lower stellar mass, when we calculate the SFR by correcting the $L_{\mathrm{FUV}}$ using the IRX-$\beta_{\mathrm{UV}}$ relation. }
  % conclusions heading (optional), leave it empty if necessary 
{}


   \maketitle
%
%________________________________________________________________

\section{INTRODUCTION}\label{introduction}

Understanding the formation mechanisms and evolution of galaxies with cosmic time has been, and is still, one of the major goals of both observational and theoretical astronomy. High redshift galaxies play a key role in developing our knowledge about the evolution  of  galaxies  and  intergalactic medium (IGM) in the distant universe. Since such galaxies are being studied within a few Gyrs after the Big Bang, they provide a unique probe of the physics of one of the first generations of stars. One of the most well-known and efficient techniques for detecting galaxies at redshifts $z > 2.5$ is the Lyman-break or dropout technique \citep{steidel96}. This technique selects galaxies over a certain redshift interval by combining the photometry in three broad-bands to detect the Lyman-break, which is the drop of the ultraviolet (UV) flux observed in correspondence with the absorption of the intergalactic and interstellar medium at $\lambda < 912$ $\AA$ (Lyman limit, the wavelength below which the ground state of neutral hydrogen may be ionized). 

The LBGs represent the largest sample of star-forming galaxies known at high redshifts ($z > 2.5$), due to the efficiency of their selection technique. They form a key population to investigate the mass assembly of galaxies during the first Gyrs of the Universe. Since the mid-1990s, their optica/near-infrared (NIR) rest-frame spectrum, combining both photometry and spectroscopy, have been studied by a large number of authors providing good measurements of their stellar mass, $10^{9-10}$ M$_{\sun}$, and star-formation rate, 10-100~M$_{\sun}$~yr$^{-1}$ \citep{Shapley01, Madau1996, steidel96, Giavalisco2002, Blaizot2004, Shapley2005, Verma2007, Magdis2008, Stark2009, ChapmanCasey2009, LoFaro2009, Magdis2010a, Pentericci2010, Oteo2013a, Bian2013}. Giving their masses and typical star-formation rate (SFR) in the context of the $\Lambda$CDM model, the LBGs are believed to be the building blocks from which, by merging processes, a fraction of massive local galaxies (L > L$^{*}$) have formed \citep{Somerville01, Baugh05}. The LBGs with the highest SFR ($> 100M_{\odot} yr^{ -1 }$) are thought to be progenitors of the present-day elliptical galaxies and the passive red galaxies at $z \sim 2-3$ \citep{Verma2007, Stark2009}.

The most commonly adopted LBG SFR tracer is from the rest-frame UV, optical and NIR at high redshift, where most of the energy is emitted by young stars (ages around 10 to 100 Myr). However, interstellar dust scatters and/or absorbs the light emitted by young stars, hence only a fraction of the energy output from star-formation is observed in the UV and the rest is re-emitted over the full IR range 8-1000 $\mu$m. \cite{Burgarella2013} showed that, even at $z = 3.6$, about half of the star formation still resides in the far-IR. So, it is necessary to combine both these tracers to determine the complete energy budget of star formation.   However, statistical and representative information on the dust emission of these LBGs are still missing, due to their faintness in the far-IR and sub-mm wavelength, which is very likely related to their low dust content \citep{Giavalisco2002}.

Only a few LBGs, selected using the dropout technique, have been directly detected at $z\sim3$ in the mid-infrared (mid-IR)  \citep{Magdis2010a}, far-IR \citep{Oteo2013,Magdis2012, Casey2012} and sub-mm \citep{Capak2015,ChapmanCasey2009, Chapman2000}, this could be related to the fact that the most dust obscured objects may be located out of the colour-colour selection and also because of their intrinsic faintness in the IR. This sub-sample of LBGs that was detected in the far-IR and sub-mm is extremely biased and not representative of the LBG population in terms of stellar mass, dust attenuation and SFR. They are more likely the equivalent of sub-millimeter bright galaxies, as suggested in \cite{Burgarella2011}. Previous works have succeed to detect them \citep{Magdis2010b, Magdis2010c, Rigopoulou2010,Davies2013} using stacking analysis in the far-IR using (922, 68 and 48) LBGs, respectively. The stacking analysis combines the signal of multiple sources that have been previously selected in other wavelength observations \citep{Dole2006,Marsden2009,Bethermin2012,Heinis2013}. However, these above samples contain a low number of objects that make it difficult to obtain a representative sampling of the LBGs characteristics. The sample used by \cite{Magdis2010b}, \cite{Magdis2010c} and \cite{Rigopoulou2010} is IRAC and MIPS detected LBGs. They are clearly biased towards the IR-bright, massive and/or dusty LBGs. The most recent results from \cite{Coppin2015}, presented the stacking of LBGs at $z\sim3$, 4 and 5 in 850~$\mu$m images (with SCUBA-2), where they enlarged the sample of LBGs studied up to 4201 at $z\sim3$.

The determination of the dust attenuation and star-formation rate using the UV+IR emission is very challenging due to the lack of deep IR data. Most of the studies of LBGs at high-$z$ use empirical recipes to correct the UV emission for dust attenuation. The most well-known is the relation between $\beta_{\mathrm{UV}}$ and the IR to UV luminosity ratio by \citeauthor{Meurer1999} (M99, hereafter 1999), which is calibrated on local starburst galaxies. The $\beta_{\mathrm{UV}}$ and the L$_{FUV}$ can be easily derived using the rest-frame UV colors when optical/NIR photometry data are available at high-$z$ and then one can compute the corrected UV luminosity using the M99 relation. This method has uncertainties coming from different aspects: the $\beta_{\mathrm{UV}}$ value is sensitive to the intrinsic UV spectral slope of galaxies which depends on the metallicity, age of the stellar population and star formation history; the relation between the dust attenuation and $\beta_{\mathrm{UV}}$ depends on the dust properties and geometry \citep{Calzetti2001}. The M99 relation is derived from local starburst galaxies and might not be valid for more ''normal'' star-forming galaxies \citep{Buat2005}. Indeed, various recent studies of local star-forming galaxies and high-redshift UV-selected galaxies have found that the relation between dust attenuation and $\beta_{\mathrm{UV}}$ does not follow the M99 relation \citep{Casey2015, Capak2015}. We might wonder, however, if LBGs at low and high redshifts do.

It is also necessary to investigate the link between dust attenuation and other galaxy properties apart from $\beta_{\mathrm{UV}}$ in order to be able to correct for dust attenuation in a statistical on data collected in surveys. On the one hand, we know that the UV luminosity is not well correlated with dust attenuation \citep{Heinis2013,Heinis2014}. On the other hand, the stellar mass shows a good correlation with dust attenuation and does not appear to evolve with redshift \citep{Heinis2014, Pannella2009, Ibar2013}.

The present project makes use of the COSMOS field \citep{Scoville2007} that covers 1.4 $\times$ 1.4 sq. deg. to select a large LBG sample (about 22,~000 objects) and stack them in the far-IR and sub-mm wavelength. The size of our sample allows us to divide them in several bins of L$_{\mathrm{FUV}}$, $\beta_{\mathrm{UV}}$ and $M_{*}$ to present the observed statistics of the dust properties of LBGs at $z \sim 3$, as witnessed by the {\sl Herschel} Space Observatory from {\sl Herschel}-PACS+SPIRE data\footnote{From two {\sl Herschel} Large Programs: PACS Evolutionary Probe \citep{Lutz2011} and the {\sl Herschel} Multi-tiered Extragalactic Survey \citep{Oliver2012}} and AzTEC from the Atacama Sub-millimetre Telescope Experiment (ASTE).

This paper is organized as follows. In Sect.~2, we present how we select the LBGs and gather the data to further analyse our sample. In Sect.~3, we detail the methodology used to stack the LBGs in the far-IR and sub-mm, including incompleteness and clustering corrections applied to obtain valid results. In Sect.~4, we present the dust properties of our LBGs and discuss them in the context of their formation and evolution. Finally, we present our conclusions in Sect.~5. Throughout this paper, we use a standard cosmology with $\Omega_{\mathrm m}$=0.3, $\Omega_{\Lambda}$=0.7 and $H_0$= 70 Km s$^{-1}$ and AB magnitude system. We employ the \cite{Chabrier2003} initial mass function (IMF). When comparing to other studies, we assume no conversion is needed for SFR and stellar mass estimates between \cite{Kroupa2001} and \cite{Chabrier2003} IMFs. When converting from \cite{Salpeter1955} IMF to \cite{Chabrier2003} IMF, we divide $M_{*}$ $_{\mathrm Salpeter}$ by 1.74 \citep{Ilbert2010}, and SFR$_{\mathrm Salpeter}$ by 1.58 \citep{Salim2007}.

%__________________________________________________________________

\section{DATA}\label{data_sample}
%The COSMOS survey is designed to probe the evolution of galaxies and active galactic nuclei (AGNs) up to z$\sim$6 over a large enough sky region, $\sim$2 deg$^{2}$, to address the role of environment and large scale structures and star formation rate (SFR). It is based on deep multi-wavelength observations from X-ray to radio (e.g. Zamojski et al. 2007; Hasinger et al. 2007; Taniguchi et al. 2007; Capak et al. 2007; Sanders et al. 2007; Bertoldi et al. 2007; Schinnerer et al. 2007; Elvis et al. 2009; McCracken et al. 2010; Lutz et al. 2011; Oliver et al. 2012).  
\textbf{\subsection{UV/Optical/NIR data}\label{data_optical}}

We use optical imaging from the COSMOS field \citep{Capak2007, Taniguchi2007}, more specifically, $V_{\mathrm J}$ and $i^{+}$ bands. The images in the $V_{\mathrm J}$ and $i^{+}$ were obtained from the SUBARU telescope using the Suprime-Cam instrument. They cover the entire COSMOS field reaching a 5$\sigma$ depth of 26.5 and 26.1, respectively, for a 3$^{\prime\prime}$ aperture. 

In addition, we use the multi-color catalogue (Capak et al. 2007, version 2.0). The fluxes are measured in different bands from data taken at the Subaru (broad-bands; $B_{\mathrm J}$, $V_{\mathrm J}$, $g^{+}$, $r^{+}$, $i^{+}$, $z^{+}$,  intermediate-bands; IA427, IA464, IA484, IA505, IA527, IA574, IA624, IA679, IA709, IA738, IA767, IA827 and narrow-bands; NB711, NB816), CFHT ($u^{*}$, $i^{*}$, $H$ and $K_{s}$ bands), UKIRT ($J$-band), UltraVISTA ($Y$,$ J$, $H$ and $K_{s}$), Spitzer (3.6 - 8 $\mu$m) and $GALEX$ (1500 - 3000 \AA). The photometry has been performed using SExtractor in dual-image mode, where the source detection has been run on the deepest image, $i^{+}$. For the UV-NIR data, the point spread function (PSF) varies in the range between 0.5$^{\prime\prime}$ and 1.5$^{\prime\prime}$ from the $u^{*}$ to the $K$ images. In order to obtain accurate colors, all the images have been degraded to the same PSF of 1.5$^{\prime\prime}$ following the method described in \cite{Capak2007}. The final photometric catalogue containes PSF-matched photometry for all the bands, measured over an aperture of 3$^{\prime\prime}$ diameter at the position of the $i^{+}$ band detection. It also provides the aperture correction to calculate the total flux for each object in the field.

\textbf{ 
\subsection{Far-infrared data}\label{farinfrared}}

We use observations of the COSMOS field from the ESA Herschel Space Observatory \citep{Pilbratt2010}. The \textit{PACS}, (Photodetector Array Camera and Spectrometer,  \citealt{Poglitsch2010}) Evolutionary Probe survey (\textit{PEP}, \citealt{Lutz2011}) mapped the COSMOS field at 100 and 160~$\mu$m with a point-source sensitivities of 1.5 mJy and 3.3 mJy and a PSF full width half maximum (FWHM) of 6.8$^{\prime\prime}$ and 11$^{\prime\prime}$. Also, the Spectral and Photometric Imaging Receiver (SPIRE,\citealt{Griffin2010}) as a part of the \textit{HerMES} (\textit{Herschel Multi-Tiered Extragalactic Survey}, \citealt{Oliver2012}) has observed the COSMOS field at 250, 350 and 500~$\mu$m. These maps have been downloaded from HeDaM\footnote{Herschel Database in Marseille: \url{http://hedam.lam.fr/HerMES/}}. For the SPIRE maps, the PSF FWHM is 18.2$^{\prime\prime}$, 24.9$^{\prime\prime}$ and 36.3$^{\prime\prime}$, the 1$\sigma$ instrumental noise is 1.6, 1.3 and 1.9 mJy~beam$^{-1}$, and the 1$\sigma$ confusion noise is 5.8, 6.3 and 6.8 mJy~beam$^{-1}$ \citep{Nguyen2010} at 250, 350, 500~$\mu$m, respectively.

\textbf{
\subsection{AzTEC data}\label{aztec}}

The AzTEC observations in the COSMOS field have been described in \cite{Aretxaga2011}. Data reduction has been performed using the standard AzTEC pipeline (see \citealt{Scott2008,Downes2012}). The observations have been taken using the Atacama sub-millimetre Telescope Experiment (ASTE; \citealt{Ezawa2004}) to a depth $\sim$1.3 mJy beam$^{-1}$. The FWHM of these observations is set to 33$^{\prime\prime}$ by fitting the post-filtered PSFs of each set of observations with a Gaussian. The AzTEC observation covers only 0.72 deg$^{2}$ of the COSMOS field.

\textbf{
\subsection{Photometric redshift and stellar masses}\label{photometricdata}}

We use the photometric redshifts (photo-$z$) and stellar masses computed for the COSMOS field by \citeauthor{Ilbert2009} (2009, version 2.0) for i-band detected sources. The photometric redshifts in the range $1.5 < z < 4$ have been tested against the zCOSMOS faint sample and faint DEIMOS spectra, showing that the accuracy of the photo-$z$ is around 3\%  \citeauthor{Ilbert2009} (2009, version 2.0). The stellar masses have been derived from spectral energy distribution (SED) fitting to the available optical and near-infrared photometry, assuming \cite{BruzualCharlot2003} single stellar population templates, an exponentially declining star-formation history, and the \cite{Chabrier2003} IMF. The X-ray detected active galactic nuclei (AGN) have been removed from the catalogue.

This catalogue provides us with three different calculations of the photo-$z$. In this paper, we will make use of the photo-$z$ derived from the median of the probability distribution function (PDF-$z$) or the one that minimizes the $\chi^{2}$ (Chi-$z$), for more information see \citeauthor{Ilbert2009} (2009, version 2.0). You can also see Sect. \ref{lbgselection} to follow the discussion of the photo-$z$ selection for LBGs sample. 

\textbf{
\section{SAMPLE}\label{lbgsample}}

This section presents the characterization and selection of the LBG sample. First, we will present the selection of the LBG sample in the colour-colour diagram (Sect.~\ref{lbgselection}). Then, we explain how the calculation of the $L_{\mathrm{FUV}}$ and $\beta_{\mathrm{UV}}$ parameters are computed in the paper (Sect. \ref{uvluminosityslope}). Next, we build a mock catalogue that will be used to define colour-colour selection criteria for our LBGs and to characterize the completeness as a function of different parameters (Sect.~\ref{mock_cat}). Finally, we calculate the UV luminosity function (LF) from our LBGs (Sect.\ref{lumfunc}). 

\textbf{
\subsection{LBG selection}\label{lbgselection}}

Our LBGs at $z\sim3$ are selected by means of the classical U-dropout technique \citep{steidel96} using band filters $u^{*}$, $V_{\mathrm J}$ and $i^{+}$. The LBGs at $z\sim3$ must present lower fluxes in the bluest band (U-band), where the lyman break is located, since the selection requires detections in $V_{\mathrm J}$ and $i^{+}$ bands. 

We select objects brighter than the magnitudes $V_{J}$=26.5 and $i^{+}$=26.1 according to a 5$\sigma$ depth in AB mag \citep{Capak2007}. We also require our objects to have photometric redshifts included in the catalogue of \citeauthor{Ilbert2009} (2009, version 2.0), which means that the objects are out of the masked areas, and are classified as galaxies. The flux in the $u$-band should be low due to the fact that it is affected by the Lyman break, therefore part of the LBGs are not detected in this band. So,  we assign a magnitude equal to 1$\sigma$ (28.7 mag) to these objects before applying the colour selection. By using the mock catalogue (see Sect. \ref{mock_cat}) and photo-$z$, we derive the following colour selection for LBGs at $2.5 < z < 3.5$:
\begin{eqnarray}
\label{eq_criterion}
u - V_{\mathrm J}  > 1 \nonumber \\
V_{\mathrm J} - i^{+} < 0.8  \\
u - V_{\mathrm J}  > 3.2  (V_{\mathrm J} - i^{+}) \nonumber
\end{eqnarray}

  \begin{figure*}
   \centering
   \includegraphics[width=\hsize]{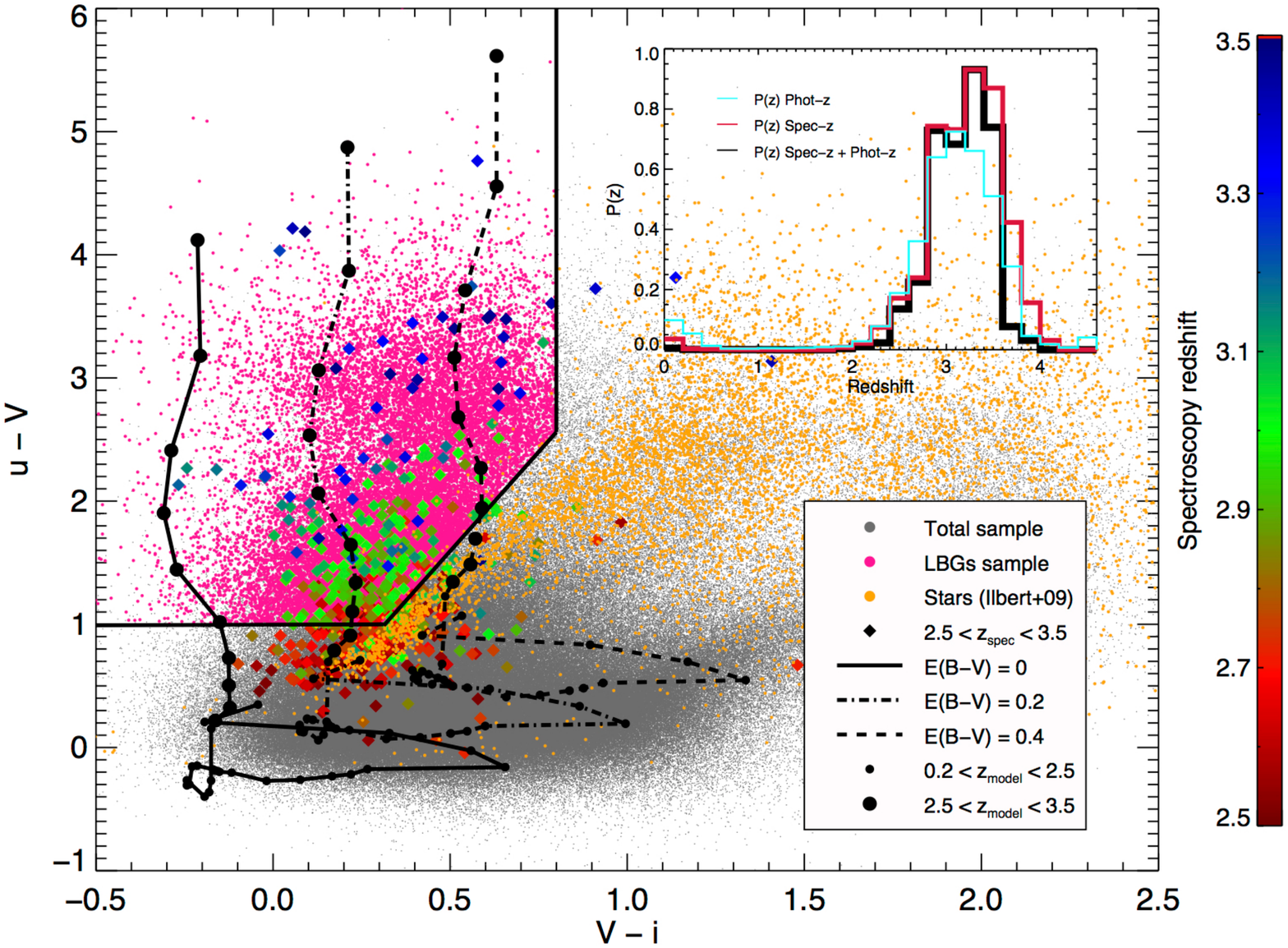}
      \caption{Selection of the LBGs in the colour-colour diagram. The small grey dots are the sample of objects detected in the COSMOS field, which have magnitudes ($V_{\mathrm J}$<26.5 and $i^{+}$<26.1) and photo-$z$ calculated from \citeauthor{Ilbert2009} (2009, version 2.0).  The pink dots are our sample selected within the limits of Eq.~\ref{eq_criterion} (black lines), with photometric redshift between 2.5 and 3.5 and log($L_{\mathrm{FUV}}$[$L_{\odot}$]) > 10.2. The yellow dots are the stars identified by \citeauthor{Ilbert2009} (2009, version 2.0). The red to blue diamonds correspond to the objects with spectroscopic redshift identify in the range 2.5 < $z_{\mathrm spec}$ < 3.5, colour-coded as a function of their $z_{\mathrm spec}$.  The black lines correspond to the expected colours in redshift evolution (small dots, $0.2 \le z \le 2.5$; large dots, $2.5 \le z \le 3.5$) for three fiducial star-forming galaxies (see Sect.\ref{modelgalaxies}) with $E(B-V)$=0, 0.2 and 0.4, respectively. The panel inserted in the top-right corner shows the probability of finding an object in the redshift range, $2.5 \le z \le 3.5$,  for our selection. The blue and red lines are the probability of finding an object with a good photo-$z$ or spectrocopic redshift (available within the COSMOS collaboration), respectively. The back line is the probability to find an object with a spectroscopic redshift after selecting the objects with photo-$z$ in the range, $2.5 \le z \le 3.5$, as we did in our sample selection (this corresponds to our redshift probability as a function of redshift).}
         \label{ccd}
   \end{figure*}

Fig.~\ref{ccd} shows the colour-colour diagram from the objects in the COSMOS field, with spectroscopic redshifts in the range $2.5 \le z \le 3.5$ and the LBG selection defined. It also shows the redshift evolution of the expected colours ($0.2 \le z \le 3.5$) for three fiducial star-forming galaxies assuming $E(B-V)$=0, 0.2, 0.4  from our mock catalogue.   

To reduce the effect of the incompleteness in the stacking process (see Sect.~\ref{stackingbias}), we keep the objects with log( $L_{\mathrm{FUV}}$[L$_{\odot}$]) $\ge$ 10.2. This luminosity corresponds to around 75\% of the completeness for the objects detected in $V_{\mathrm J}$ and $i^{+}$ bands (See fig.~\ref{completeness}). 

In order to clean the sample of lower-$z$ interlopers, we select those galaxies whose PDF-$z$ are within $2.5 \le z \le 3.5$ (20,~819 galaxies). We also keep in our sample the objects for which the PDF-$z$ are not in the adopted redshift range, but, the value of  Chi-$z$ and the error of the PDF-$z$ are in accord with our redshift range (930 galaxies). We find that the fraction of sources that we recover using our color selection with respect to the photo-$z$ is around 80~\%. We also tested our sample against the COSMOS spectroscopic master catalogue (available within the COSMOS collaboration), which contain around 500 objects within $2.5 \le z_{spec} \le 3.5$ and the 97\% of the sources present a magnitude lower than 25 in the i-band. We recovered 82~\% of the sources spectroscopically confirmed to be lying in the redshift range $2.5 \le z \le 3.5$, after selecting objects with photo-$z$ in the redshift range $2.5 \le z \le 3.5$ (as we do for our LBG selection). The inset in the Fig.~\ref{ccd} shows the probability as a function of redshift for our selection criteria. This probability is calculated in relation to the total catalogue (number of objects inside our selection criterion for each redshift bin, divided by the number of objects in the total catalogue for the same redshift bin).

The final sample of LBGs contains $\sim$22,~000 LBGs with a $z_{mean} = 3.02 \pm 0.25$. We stress that the AzTEC observations  cover 0.72 deg$^{2}$, therefore, the sample used to stack in the AzTEC data is reduced to a subsample of approximately 7,~700 LBGs.  The source selection criteria being exactly the same inside and outside the covered area, this should not introduce any bias.

The main goal of this work is to characterize the dust properties of LBGs at redshift $z\sim3$ as a function of different physical parameters. The relatively large number of LBGs included in the sample allows us to study the LBGs in several bins as a function of their $L_{\mathrm{FUV}}$, $\beta_{\mathrm{UV}}$ and $M_{*}$ to better investigate their variety. We split the sample as a function of $L_{\mathrm{FUV}}$, $\beta_{\mathrm{UV}}$ and $M_{*}$, defining size bins of 0.3 dex, 0.4 and 0.25 dex, respectively (See Table \ref{t1} for the number of the objects and interval values for each bin selected).

\textbf{
\subsection{Far-UV luminosity and slope of the UV continuum}\label{uvluminosityslope}}

$\beta_{\mathrm{UV}}$ and $L_{\mathrm{FUV}}$ along this paper are computed using the broad bands ($B_{\mathrm J}$, $V_{\mathrm J}$, $g^{+}$, $r^{+}$, $i^{+}$), intermediate bands (IA464, IA484, IA505, IA527, IA574, IA624, IA679, IA709, IA738, IA767, IA827) and narrow bands (NB711, NB816) from Capak's et al. (2007, version 2.0) catalogue. We consider the UV rest-frame wavelength range, 1250 $\AA$ < $\lambda$ < 2000 $\AA$ \citep{Calzetti1994}, to calculate $\beta_{\mathrm{UV}}$ and $L_{\mathrm{FUV}}$. We exclude the range 2000 $\AA$ < $\lambda$ < 2600 $\AA$ for two reasons, to omit the relevant dust feature at 2175 $\AA$ and to have an homogeneous rest-frame wavelengths independently of the redshift of the galaxy. We impose that the bands used in the analysis must be detected with a signal-to-noise (SNR) > 3$\sigma$. $\beta_{\mathrm{UV}}$ and $L_{\mathrm{FUV}}$ are obtained by fitting the photometry to a simple power-law SED, $f_{\lambda} \propto \lambda^{\beta_{\mathrm{UV}}} $. The $L_{\mathrm{FUV}}$ is calculated at 1600 $\AA$.

 If we consider the rest-frame wavelength range used in this work, there are at least 11 bands available to calculate $\beta_{\mathrm{UV}}$ and $L_{\mathrm{FUV}}$. The large number of bands reduces the error in the $\beta_{\mathrm{UV}}$ and $L_{\mathrm{FUV}}$ determination. The uncertainty in the photometric redshift should influence our $\beta_{\mathrm{UV}}$ and $L_{\mathrm{FUV}}$ calculations. We consider that the UV spectrum follows a simple power-law. If the photo-$z$ are perturbed according to their uncertainties, the slope of the linear fit has to remain the same. For our LBG sample (see Sect. \ref{lbgselection}), we find $\langle \sigma_{\beta_{UV}}\rangle$ = 0.3 $\pm$ 0.1. On the other hand, the uncertainty in the photometric redshift produces an influence on the $L_{\mathrm{FUV}}$ value. However, this influence is not taken into account because it is smaller than the uncertainty in our $L_{\mathrm{FUV}}$ calculation (relative error equal to 20 $\pm$ 10 \%). 

We would like to stress that our analysis is based on a statistical study of LBGs, where we split the sample in different bins of $\beta_{\mathrm{UV}}$, $L_{\mathrm{FUV}}$ and $M_{*}$. The errors showed before, for the $\beta_{\mathrm{UV}}$ and $L_{\mathrm{FUV}}$ calculations, are smaller than the bin sizes used in the binning of the sample. Therefore, they will not have a strong effect in the final results of our stacking analysis.     

\textbf{
\subsection{Mock catalogue}\label{mock_cat}}

A mock catalogue is used to characterize the completeness of the LBG sample in different parts of this work. First, the photometric catalogue will be cut at fixed $L_{\mathrm{FUV}}$ to reduce the incompleteness of the final LBGs sample (see Sect.~\ref{lbgselection}). Then, we will characterize the completeness of the LBG sample as a function of $L_{\mathrm{FUV}}$ and redshift to compute the UV LF (see Sect.~\ref{lumfunc}). Finally, we will estimate the incompleteness correction for each bin in the stacking as a function of $L_{\mathrm{FUV}}$, $\beta_{\mathrm{UV}}$ and $M_{*}$ (see Sect.~\ref{stackingbias}). Additionally, the mock catalogue will be used to define the color selection criteria for our LBGs at $z\sim3$  (see Sect.~\ref{lbgselection}). \\

\subsubsection{Model galaxies}\label{modelgalaxies}

We set up our fiducial galaxy SED model using CIGALE (\citealt{Burgarella2005, Noll2009b, Ciesla2015}; Burgarella et al.~in prep; Boquien et al.~in prep.)\footnote{Code Investigating GALaxy Emission (CIGALE), \url {http://cigale.lam.fr}}. We simulate star-forming galaxies by considering  a constant star-formation history (SFH), an age of 100 Myr (as suggested by \citealt{vanderBurg2010}),  a sub-solar metallicity (0.2 Z$_{\odot}$ as in \citealt{Castellano2014}), \cite{BruzualCharlot2003} stellar population libraries, emission lines, and a \cite{Chabrier2003} IMF. Computed template SEDs are then reddened by means of the ``standard'' \cite{Calzetti2000} attenuation law for starburst, with a Gaussian distribution in $E(B-V)$  following $<E(B-V)>~= 0.2 \pm 0.1$ and $E(B-V) \ge 0$. The chosen E(B-V) distribution matches the $\beta_{\mathrm{UV}}$ distribution from our sample and the ones from our fiducial templates. We notice that the assumed distribution for the $E(B-V)$, is also in agreement with previous studies ( e.g, \citealt{Papovich2001, Reddy2008, Castellano2014}) which suggests a mean between 0.1 and 0.2.

%. Models should reproduce this behavior to the extent required by the simulations.

Spectral synthesis modeling and external multi-wavelength information also indicate that the rest-frame UV wavelength range can be described, for $z\sim2-3$ galaxies, by starbursts phases and constant star formation (e.g., \citealt{Shapley2005}). For instance, we are aware that some degeneracy exists, e.g., with respect to the age of the stellar populations and the amount of dust attenuation, that made no unique the assumptions to model the colours of LBGs at $z\sim3$. However, the chosen parameters provide representative ultraviolet colours of galaxies at $z\sim3$, like LBGs (see \citealt{vanderBurg2010, Reddy2008}).

\subsubsection{Building the mock catalogue}\label{buildingmock}

We take advantage of the Monte Carlo approach to determine the transformation between the intrinsic properties of galaxies (UV luminosity, redshift and reddening) and their observed rest-frame UV colours. We simulate our fiducial templates by assuming the same redshift distribution as that for the observed sample (i.e. $2.5 < z < 3.5$). Then 200,~000 fiducial galaxy SEDs are created using a flat distribution in redshift and the E(B-V) distribution aforementioned. CIGALE provides us with SEDs and intrinsic properties normalised to 1~$M_{\odot}$. To create the mock catalogue, we compute the $L_{\mathrm{FUV}}$ and the UV slope for the fiducial SEDs normalized to 1~$M_{\odot}$ using the same procedure exploited for our sample and we rescale them to a flat distribution in $L_{\mathrm{FUV}}$ (9 < $log(L_{FUV} [L_{\odot}])$ < 12). The Galactic extinction is added to the mock catalogue using the \cite{Cardelli1989} extinction curve. We add a stellar mass value for each object in the mock catalogue using the rest-frame UV light to mass ratio, being aware of the limitations (see next paragraph). The stellar mass is calculated by rescaling the value of the stellar mass, given by CIGALE for the models normalized to 1~$M_{\odot}$, by the same factor used to obtain the $L_{\mathrm{FUV}}$. We also add a  gaussian scatter with a sigma equal to 0.5 dex to the stellar mass calculation.

The emission of our galaxies (modeled by assuming both a single constant SFH and one stellar population with a given distribution in $E(B-V)$) allows us to predict the UV rest-frame colours of $z\sim3$ LBGs ( e.g. \citealt{Shapley2005}), but not their stellar masses.  \cite{Sawicki2012} found for BX galaxies at $z\sim$2.3 a stellar mass  - $L_{\mathrm{FUV}}$ relation with a scatter of about 0.5 dex. Also, \cite{Hathi2013} found for LBGs at $z\sim$1-5 a relation with a scatter of about 0.3 dex. For this, we chose to include a 0.5dex dispersion in the calculation of the stellar mass to break the degeneracy in the method. We stress that we only wish to obtain a realistic stellar mass scale to see the impact to the stacking analysis (see Sect. \ref{stackingbias}).

It is well known that the $K$-band rest-frame magnitude provides a good correlation with the stellar mass \citep{Kauffmann1998}. When we created fiducial galaxy SEDs with CIGALE, we also included the 8 $\mu m$ IRAC band ($K$-band rest-frame at $z~\sim3$). We calculate the stellar mass using the mass-to-light ratio given by \cite{Magdis2010a}, including they uncertainty in their relation. The stellar mass distributions of the mock catalogue estimated from the two methods are found to be consistent. This have been done only to confirm that the first method gives a realistic stellar mass scale.

\begin{figure}[h]
   \centering
   \includegraphics[width=\hsize]{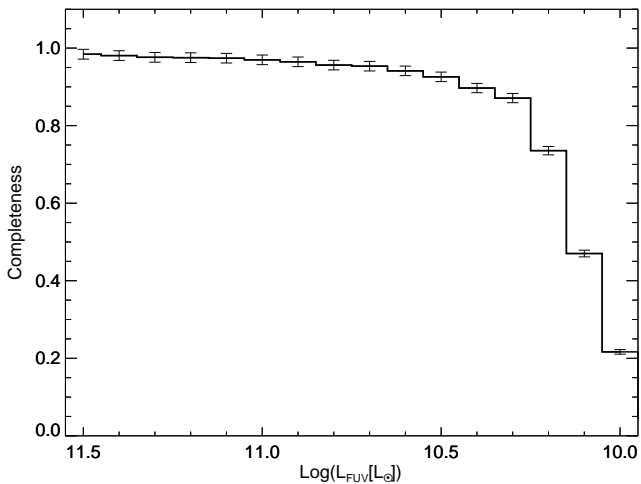}
      \caption{Completeness as a function of $L_{\mathrm{FUV}}$. We use the simulation form Sect. \ref{modelgalaxies} to compute the completeness of our catalogue as a function of $L_{\mathrm{FUV}}$ for objects detected in $V_{\mathrm J}$ and $i ^{+}$ bands in the SUBARU observations. The error bars are calculated from the Poisson noise term.}   
         \label{completeness_image}
   \end{figure}
   
\subsubsection{Completeness}\label{completeness}

To quantify the completeness as a function of different parameters ($L_{\mathrm{FUV}}$, $\beta_{\mathrm{UV}}$, $M_{*}$ and redshift), which we will use in our sample selection, UV LF calculation, and stacking analysis, we insert the objects from the mock catalogue into $i^{+}$ and $V_{\mathrm J}$ images. These two images are the ones where we perform the selection of our LBG sample (see Sect.\ref{lbgselection}). LBGs have typical half-light radii of $r \sim 0.1^{\prime\prime}-0.3^{\prime\prime}$ \citep{Giavalisco2002} and thus are unresolved by our observations. They can be treated as point sources.  Futhermore, we assume a Gaussian profile in agreement with the PSF of the images to inject the objects. We inject 10,~000 simulated objects, each time, in random position on the image. We then attempt to recover these simulated galaxies using the same procedure as in \cite{Capak2007}. We repeat it for the 200,~000 objects of the mock catalog. Fig. \ref{completeness_image} shows the completeness as a function of the $L_{\mathrm{FUV}}$ for objects detected in $i^{+}$ and $V_{\mathrm J}$ images.

\textbf{
\subsection{Luminosity function}\label{lumfunc}}

We derive here the UV luminosity function (LF) to test whether our LBG sample is so far representative of the LBG population obtained in previous works. The UV LF is calculated by using the $V_{\mathrm max}$ method \citep{Schmidt1968} for our total LBG sample. The effective volumes ($V_{\mathrm max}$) of our survey are given by:
\begin{equation}
\phi^{i}~(M_{\mathrm{UV}})dM=\sum_{i}^{N_{M_{\mathrm{UV}}^{i}<M_{\mathrm{UV}}<M_{\mathrm{UV}}^{i+1}}}~\dfrac{p(M_{\mathrm{UV}}^{i}, z)}{V_{tot}^{i}};\\
V_{\mathrm tot} =A \int \frac{dV_{\mathrm c}}{dz} dz 
\end{equation}

where $N_{M_{\mathrm{UV}}^{i}<M_{\mathrm{UV}}<M_{\mathrm{UV}}^{i+1}}$ is the number of objects in the bin of $M_{\mathrm{UV}}$, $A$ is the field area in deg$^{2}$ and $\frac{dV_{\mathrm c}}{dz}$ is the comoving volume per deg$^{2}$. The $p(M_{\mathrm{UV}}^{i}, z)$ is a function calculated using the mock catalogue and the completeness (see Sect. \ref{mock_cat}). This corresponds to the number of sources recovered with an observed magnitude in the interval [$M_{\mathrm{UV}}$; $M_{\mathrm{UV}}$+$\Delta M$], that are selected as dropouts and divided by the number of injected sources with an intrinsic magnitude in the same interval [$M_{\mathrm{UV}}$; $M_{\mathrm{UV}}$+$\Delta M$] and a redshift in the interval [$z$; $z$+$\Delta z$].The magnitude is measured at rest-frame wavelength 1600~$\AA$. The resulting luminosity function is binned to $\Delta$mag=0.25.

The uncertainties in the luminosity function is derived combining the Poisson noise term, cosmic variance and the determination of $p(m, z)$. The cosmic variance in the COSMOS field using the mean mass of our sample and a redshift bin size, $\Delta z=1$, amounts to 5\% of LF \citep{Moster2011}. In the calculation of $p(m,z)$ we estimate an uncertainty of around 5\%, this are coming from the poisson noise in the determination of the completeness in $M_{\mathrm{FUV}}$ and $z$. We add the three uncertainties to compute the error on the LF.

  \begin{figure}[h]
   \centering
   \includegraphics[width=\hsize]{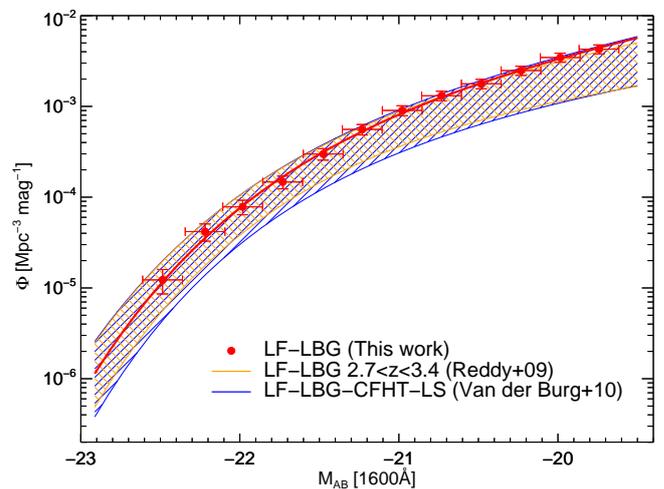}
      \caption{UV LF for our LBGs sample in the redshift bin, $2.5 < z <3.5$. Our data points and best-fit Schechter functions are shown in red. We compare them with the 1$\sigma$ region for the UV LF of LBGs calculated by \citeauthor{Reddy2009} (yellow, 2009) and \citeauthor{vanderBurg2010} (blue, 2010) in the redshift bin, $2.7 < z < 3.4$.}
         \label{lum_func}
   \end{figure}

We fit a Schechter function \citep{Schechter1976} to the binned data points,
\begin{equation} 
\phi(M)~dM = 0.4 ln(10) \phi^{*} 10^{0.4(\alpha + 1)(M^{*}-M)} exp(-10^{0.4(M^{*}-M)})
\end{equation}
with $M^{*}$ being the characteristic magnitude, $\alpha$ the faint-end slope, and $\phi^{*}$ the overall normalization. We find $\phi^{*}$ = (2.65 $\pm$ 0.41) 10$^{-3}$ , $M^{*}$ = -20.94 $\pm$ 0.08 and $\alpha$ = -1.84 $\pm$ 0.09. As can be seen in Fig. \ref{lum_func}, our results agree within the 1$\sigma$ region of the UV LF at $2.7 < z < 3.4$ by \cite{Reddy2009} and \cite{vanderBurg2010}, but we found the UV LF systematically higher than these previous results, and that this excess increases toward the faint-end. In Sect. \ref{lbgselection}, we find that the fraction of sources recovered, combining both the color and the photo-$z$ selections, with respect to the spectroscopic sample is around 82~\% (which means that we are including around 20\% of sources that have a photo-$z$ within the redshift range, but the spectroscopic redshifts are out of the target redshift range). The uncertainty in the photo-$z$ calculation also increases for fainter objects. This can be one of the possible explanations for the over density of objects observed in our LF.  However, we should take into account that some differences can be also found due to differences in the selection of the sample and purity corrections. We are not using the same filters and colour-selection to define our LBG sample than the previous works (UBr bands), this can generate differences in the distribution of redshift between the different LBG samples. The previous LFs are computed in a shorter redshift range ($2.7< z < 3.4$). Note that we clean the sample by using a photo-$z$ selection (see Sect. \ref{lbgselection}). Nevertheless the other mentioned LFs use simulations to obtain the fraction of stars and low redshift interlopers that contaminate the sample. The cosmic variance between the different fields can also produce differences in the LFs. Despite all of this, our $M^{*}$ and $\alpha$ estimations of the LF are found to be within 1$\sigma$, and $\phi^{*}$ within 1.5$\sigma$ with respect to \cite{Reddy2009} estimates. Therefore, We conclude that our LBG sample is roughly representative of the LBG population.

\section{STACKING MEASUREMENTS}\label{stackinganalisis}

The stacking is a technique that combines the signal from multiple sources selected from observations at other wavelengths (e.g \citealt{Dole2006,Marsden2009,Bethermin2012,Heinis2013}). With this stacking technique we are able to obtain a statistically significant measure of physical parameters in faint galaxies at high redshift which are otherwise lost beneath the noise levels, at the expense of averaging over individual properties. We simultaneously stack $30\times30$ pixel cut-outs using the IAS library (Bavouzet 2008 and \citealt{Bethermin2010b})\footnote{\url{http://www.ias.u-psud.fr/irgalaxies/downloads.php}} in PACS (100 and 160~$\mu$m images), SPIRE (250, 350 and 500~$\mu$m images) and the AzTEC (1.1 mm image).

\textbf{ 
\subsection{Stacking corrections}}

LBGs are clustered with other LBGs and other populations of star-forming galaxies at high redshift \citep{Hickox2012}, causing a non homogeneous background in the stacked image (Bavouzet 2008; \citealt{Bethermin2010b}; \citealt{Heinis2013}). To get valid and reliable results, we have to correct our measurements for two effects: incompleteness and clustering of the sample. \\

\subsubsection{Correcting for incompleteness of the input catalogue in the dense regions}\label{stackingbias}

A bias is produced in the stacked image when the population of sources is not complete (\citealt{Dole2006}; Bavouzet 2008; \citealt{Bethermin2010b}; \citealt{Heinis2013}; \citealt{Viero2014}). During the detection process in the optical images, we miss some of the faint objects located in the dense areas or close to bright objects. Therefore, we lose the contribution of the high background areas when we stack in the far-IR images, causing a negative flux contribution near to the stacked object in relation to the global background.

We evaluate the contribution of this bias in the stacking using the method from \cite{Heinis2013}. They showed that a bias effect of this nature increases inversely proportional to $L_{\mathrm{FUV}}$. They selected the sample from the u-band, which corresponds to the UV ($\lambda_{\mathrm{UV}} = 1600~\AA$) in the rest-frame wavelength at $z\sim1.5$. Our selection sample is different, the UV rest-frame wavelength being located in the r band at $z\sim3$, and we select the objects in the $V_{\mathrm J}$ and $i^{+}$ bands. Therefore, our stacking bias effects are related to the detection of the objects in the $V_{\mathrm J}$ and $i^{+}$ images. To quantify the impact of this bias, we use the mock catalogue created to estimate the completeness (see Sect. \ref{modelgalaxies}). We split the recovered simulated sources in the $V_{\mathrm J}$ and $i^{+}$ images from the mock catalogue in the same way as our LBGs sample. To build the bias correction map in each bin, we stack in the far-IR and sub-mm images the random positions where we previously recovered the simulated sources in the $V_{\mathrm J}$ and $i^{+}$ images from the mock catalogue. In order to correct our stacking measurement for this effect, we subtract the bias maps, from the maps obtained by stacking at the position of the true LBGs. 

\begin{figure}[h]
   \centering
   \includegraphics[width=\hsize]{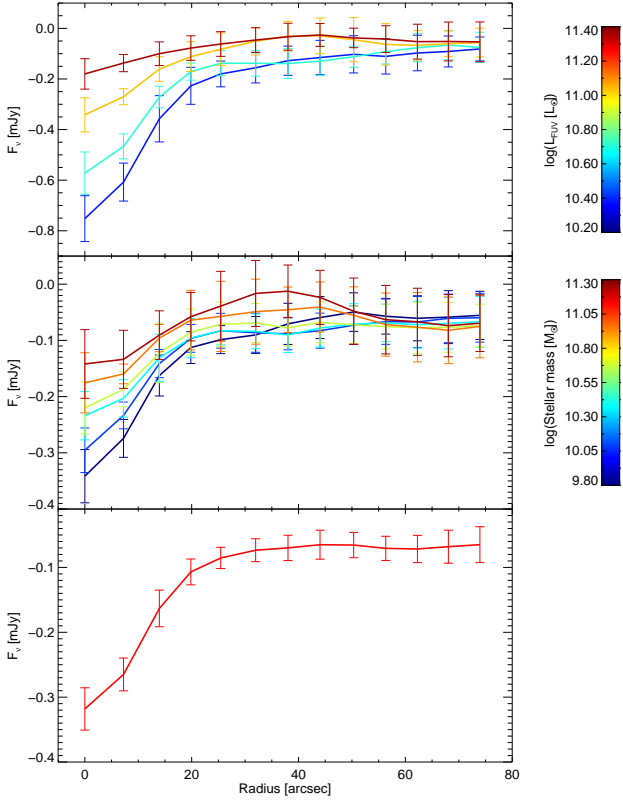}
      \caption{Radial profile of the bias correction maps used for each bin of the stacking as a function of the $L_{\mathrm{FUV}}$ (top)  and $M_{*}$ (middle) at the SPIRE 250$\mu$m. The bottom panel shows the bias correction used to correct each bin of the stacking as a function of the $\beta_{\mathrm{UV}}$, due to the fact that we did not find any difference in the bias correction as a function of the $\beta_{\mathrm{UV}}$.}
         \label{bias_correct}
   \end{figure}

Fig.~\ref{bias_correct} shows the radial profile of the bias correction maps as a function of the $L_{\mathrm{FUV}}$  and $M_{*}$. In agreement with \cite{Heinis2013}, we find that the amplitude of this effect increases for objects with fainter $L_{\mathrm{FUV}}$. This result is related to the well-known effect according to which the detection efficiency is lower in the denser of the UV images used for the colour-colour selection, and in particular for faint objects that are close to the brighter ones. This effect is also dependent on $M_{*}$, but it presents a lower difference in the amplitude between the different bins of $M_{*}$ than for the $L_{\mathrm{FUV}}$. However, we did not detect any dependence in the UV-slope. These last two effects are related to the UV-luminosity dispersion in each bin when we stack as a function of the $M_{*}$ and $\beta_{\mathrm{UV}}$. Considering that we did not find any evolution as a function of the UV slope, a bias correction made using the full recovered mock catalogue is used to correct each bin of the stacking as a function of the $\beta_{\mathrm{UV}}$ values (see bottom panel in the Fig. \ref{bias_correct}).

\subsubsection{Correcting for clustering of the input catalogue}\label{clustering_correction}

The large PACS, SPIRE and AzTEC beams can be contaminated by neighbouring sources due to the large PSF. This effect would lead to overestimating the mean flux of our sample due to the clustered nature of the sources. We use the formalism developed by Bavouzet (2008) and \cite{Bethermin2010b} to address this. They assumed that the excess probability to find another galaxy from a sample compared to a randomly distributed population is proportional to the angular auto-correlation function. The two-dimensional profile of the resulting stacking can then be written as:
\begin{equation}\label{eq_stack}
I (\theta,\phi) = \alpha \times PSF( \theta , \phi ) + \beta \times w( \theta , \phi ) \ast PSF( \theta , \phi ) + \gamma
\end{equation}

Here $I (\theta,\phi)$ is the stacked map after bias correction, $\alpha$ is the average flux of the stacked population, $PSF( \theta , \phi )$ is the point spread function (PSF) at the stacked wavelength,  $w( \theta , \phi )$ is the angular auto-correlation function of the input catalogue, $\beta$ is a parameter related to the density of the input population and $\gamma$ is the constant related to the background in the stacked image.

We measure $w(\theta,\phi)$ using the \cite{Landy1993} estimator in the position of our LBGs sample. We fit it with a power law $w( \theta , \phi ) \propto \theta^{- \delta}$, finding that the correlation function is well modeled with $\delta = 0.63$. We check the slope of the auto-correlation function in each bin as a function of different parameters and we do not find any significant change \citep{Heinis2013}. Hence, we consider only the best fit to the auto-correlation function of the full sample. The auto-correlation function diverges when $\theta=0$, therefore, we assume that it takes effect out of 3 arcsec in relation to the center of the stacked image. 

\begin{figure}[h]
   \centering
   \includegraphics[width=\hsize]{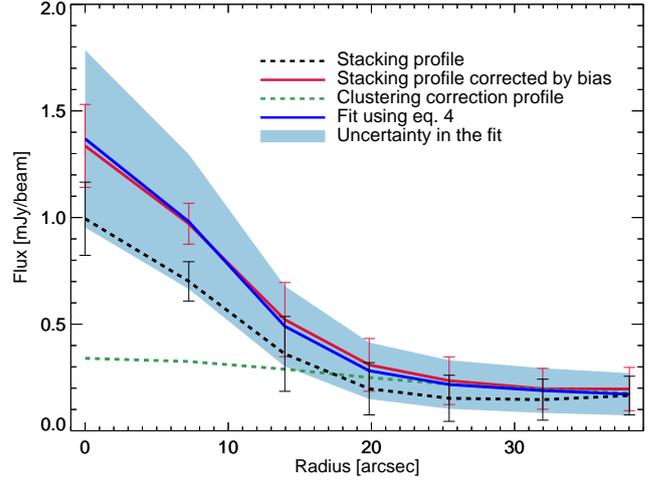}
      \caption{Example for the solution of the Eq. \ref{eq_stack} as a radial profile, where we present the contribution of the corrections for incompleteness and clustering of the stacked galaxies. We show here the results for the third bin of the stacking as a function of $L_{\mathrm{FUV}}$ in the 250~$\mu$m band (10.80 < log($L_{\mathrm{FUV}}$[L$_{\odot}$]) > 11.1, LBG-L3 see Table \ref{t1}). We obtain the parameters of the fit: $\alpha$ =0.99$\pm$0.26 , $\beta$ =0.46$\pm$0.16 and $\gamma$ =-0.12$\pm$0.08. The black dashed line is the profile of the stacked LBGs population inside of the $L_{\mathrm{FUV}}$ bin. The red line corresponds to the profile of the stacked LBGs population inside of the $L_{\mathrm{FUV}}$ bin corrected for the incompleteness of the input catalogue, I ($\theta$,$\phi$). The green dashed line is the contribution for the clustering of the input catalogue, $\beta$ $\times$ w( $\theta$ , $\phi$ ) $\ast$ PSF( $\theta $, $\phi$ ). The blue line corresponds to the sum of the real emission of the object, $\alpha$ $\times$ PSF( $\theta$ , $\phi$ ), the contribution for the clustering of the input catalogue, $\beta$ $\times$ w( $\theta$ , $\phi$ ) $\ast$ PSF( $\theta $, $\phi$) and the background, $\gamma$. The blue region shows the uncertainty of the results using bootstrap resampling. This plot is only a 1D illustration for the solution of Eq. \ref{eq_stack}, but, we have solved it in 2D to obtain the stacked images for this work.}
         \label{ajust_object}
   \end{figure}

We solve the matrix system of Eq.~\ref{eq_stack} to obtain the flux density of our stacked objects. Bootstrap resampling is used to obtain the mean values and errors. We repeat the above procedure on 3000 random bootstrap samples, and the 1$\sigma$ of the distribution of the derived fluxes is adopted as the uncertainty of our results. Fig.~\ref{ajust_object} presents an example in profile for the solution of Eq.~\ref{eq_stack} in a specific $L_{\mathrm{FUV}}$ bin for the 250 $\mu$m image and its uncertainty. We also illustrate the contribution of the incompleteness and the clustering correction from the input catalogue.

\textbf{
\subsection{Stacking results}}

\begin{figure}[h]
   \centering
   \includegraphics[width=\hsize]{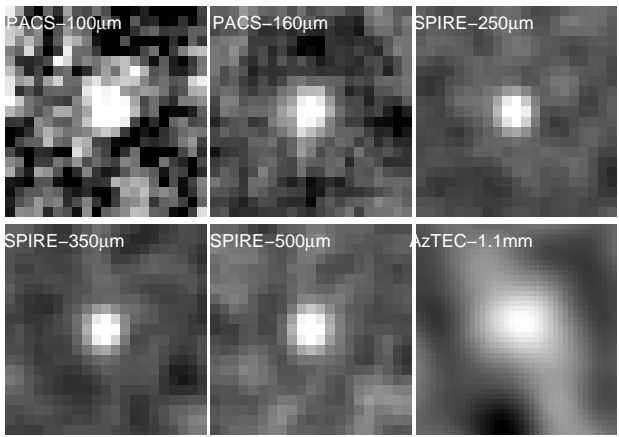}
      \caption{An example of the stacked images in the PACS, SPIRE and AzTEC bands. Here, we present the bin, LBG-M3 (see Table~\ref{t1}), of the stacking as a function of stellar mass. The pixel size here correspond to 2", 3$^{\prime\prime}$, 6$^{\prime\prime}$, 8.3$^{\prime\prime}$, 12$^{\prime\prime}$ and 3", respectively.}
         \label{stacking_example}
   \end{figure}

Fig.\ref{stacking_example} shows the stacked images for a specific mass bin in PACS, SPIRE and AzTEC bands, the stacked images have been corrected for the bias maps. It indicates a significant detection of the emission for all the PACS, SPIRE and AzTEC bands of the LBG population.

Table \ref{t1} shows the results of the stacking as a function of $L_{\mathrm{FUV}}$, $\beta_{\mathrm{UV}}$ and $M_{*}$. We present the mean values of the parameters characterizing each population bin: $L_{\mathrm{FUV}}$, $\beta_{\mathrm{UV}}$, $M_{*}$, photo-$z$, number of stacked LBGs and their respective uncertainties (estimated by the standard deviation). The fluxes from the stacking measurement in each band and their errors are also listed. We are also giving the SNR calculate using the flux of our stacked LBGs and the $\sigma$\footnote{We calculate the $\sigma$ of the staked images, $\sigma = \sqrt{\sigma_{back}^{2} + \sigma_{back-bias}^{2}}$, where $\sigma_{back}$ corresponds to the sigma of the stacking at ``X'' random position (where ``X'' is the number of LBGs in each bin) and $\sigma_{back-bias}$ corresponds to the sigma of the stacking at ``Y'' random position (where ``Y'' is the number of objects of the mock catalog stacked in each bin) and repeated both 1000 times.} of the stacked images, this are presented for each bin and band brackets. 

\begin{table*}
\caption{\label{t1}Stacking results}
\centering
\resizebox{\textwidth/1}{!}{
 \renewcommand{\arraystretch}{1.5}
\begin{tabular}{ccccccccccccc}
\hline
ID & Range & $\langle$log($L_{\mathrm{FUV}}$[L$_{\sun}$])$\rangle$ & $\langle$log($M_{*}$ [M$_{\sun}$])$\rangle$ & $\langle$ $\beta$ $\rangle$ & $\langle$z$\rangle$ & N$_{gal}$ & $S_{100}$[mJy] & $S_{160}$[mJy] &$S_{250}$[mJy] &$S_{350}$[mJy] &$S_{500}$[mJy] &$S_{1100}$[mJy] \\
\hline
\multicolumn{13}{c}{\textbf{Stacking as a function of $L_{\mathrm{FUV}}$ (LBG-$L$)}}\\
\hline

LBG-$L$1 & 10.2 - 10.5 & 10.35$\pm$0.09 & 9.67$\pm$0.47 & -1.26$\pm$0.60 & 2.98$\pm$0.23 & 13078 & 0.09$\pm$0.03 (3.4) & 0.27$\pm$0.08 (3.7) & 0.50$\pm$0.14 (5.5) & 0.46$\pm$0.15 (4.7) &  0.40$\pm$0.15 (4.2) & <0.11 \\
LBG-$L$2 & 10.5 - 10.8 & 10.63$\pm$0.09 & 9.79$\pm$0.44 & -1.26$\pm$0.54 & 3.08$\pm$0.25 & 6601 &  0.10$\pm$0.04 (3.4) & 0.35$\pm$0.10 (4.3) & 0.85$\pm$0.18 (8.3) & 0.91$\pm$0.19 (8.2) &  0.77$\pm$0.18 (7.2) & <0.15 \\
LBG-$L$3 & 10.8 - 11.1 & 10.91$\pm$0.08 & 9.96$\pm$0.38 & -1.19$\pm$0.44 & 3.11$\pm$0.25 & 1815 &  0.16$\pm$0.06 (3.4) & 0.46$\pm$0.14 (3.5) & 0.99$\pm$0.26 (5.9) & 1.18$\pm$0.29 (6.8) &  0.85$\pm$0.28 (4.9) & <0.28 \\
LBG-$L$4 & 11.1 - 11.4 & 11.20$\pm$0.08 & 10.14$\pm$0.38  & -1.18$\pm$0.40 & 3.13$\pm$0.24 & 255 &    <0.80            & 0.74$\pm$0.46 (2.4) & 3.28$\pm$0.90 (7.9) & 3.08$\pm$0.89 (6.9) &  2.78$\pm$0.81 (6.5) & <0.82 \\
\hline
\multicolumn{13}{c}{\textbf{Stacking as a function of $\beta_{\mathrm{UV}}$ (LBG-$\beta$)}}\\
\hline
LBG-$\beta$1 & -1.7 - -1.1 & 10.50$\pm$0.23 & 9.70$\pm$0.41 & -1.38$\pm$0.17 & 3.04$\pm$0.23 & 8659 & <0.063              &  <0.15              &     <0.28	             & 0.27$\pm$0.12 (3.6) & 0.39$\pm$0.13 (5.2) & <0.15  \\ 
LBG-$\beta$2 & -1.1 - -0.7 & 10.50$\pm$0.23 & 9.79$\pm$0.50 & -0.91$\pm$0.12 & 2.99$\pm$0.25 & 5268 & 0.11$\pm$0.04 (4.4) & 0.39$\pm$0.09 (5.3) & 0.78$\pm$0.16 (8.5)  & 0.74$\pm$0.17 (7.6) & 0.56$\pm$0.16 (6.0) & <0.16  \\
LBG-$\beta$3 & -0.7 - -0.3 & 10.48$\pm$0.21 & 9.87$\pm$0.56 & -0.53$\pm$0.12 & 2.95$\pm$0.27 & 2563 & 0.19$\pm$0.05 (5.2) & 0.56$\pm$0.13 (5.6) & 1.15$\pm$0.24 (8.8)  & 1.15$\pm$0.25 (8.3) & 1.01$\pm$0.25 (7.5) & <0.25 \\
LBG-$\beta$4 & -0.3 -  0.1 & 10.44$\pm$0.19 & 9.92$\pm$0.64 & -0.15$\pm$0.11 & 2.92$\pm$0.27 & 730  & <0.35               & <0.74               & 2.47$\pm$0.50 (10.2) & 2.62$\pm$0.50 (10.1)   & 2.13$\pm$0.47 (8.5) & <0.53 \\
LBG-$\beta$5 & 0.1 -  0.5 & 10.43$\pm$0.18 & 10.12$\pm$0.50 & -0.26$\pm$0.11 & 2.89$\pm$0.26 & 141  & 0.54$\pm$0.19 (3.7) & 1.91$\pm$0.61 (4.7) & 3.96$\pm$1.04 (7.3)  & 3.33$\pm$1.05 (5.6) & 1.82$\pm$0.91 (3.2) & <1.12 \\
\hline
\multicolumn{13}{c}{\textbf{Stacking as a function of stellar mass (LBG-$M$)}}\\
\hline
LBG-$M$1 & 9.75-10.00 &  10.51$\pm$0.22 &  9.87$\pm$0.08 & -1.18$\pm$0.52 & 3.04$\pm$0.26 & 5461 & <0.10         & 0.27$\pm$0.09 (3.4) &  0.49$\pm$0.15 (4.9) &  0.47$\pm$0.19 (4.7) &  0.46$\pm$0.17 (4.6) & <0.06 \\
LBG-$M$2 & 10.00-10.25 & 10.56$\pm$0.25 & 10.11$\pm$0.08 & -1.03$\pm$0.52 & 3.01$\pm$0.26 & 2811 & 0.21$\pm$0.06 (4.5) & 0.60$\pm$0.14 (6.0) &  1.38$\pm$0.25 (10.4) &  1.33$\pm$0.26 (9.5) &  0.99$\pm$0.24 (7.2) & 0.19$\pm$0.10 (1.7) \\
LBG-$M$3 & 10.25-10.50 & 10.59$\pm$0.28 & 10.35$\pm$0.08 & -0.90$\pm$0.52 & 2.98$\pm$0.26 & 1319 & 0.21$\pm$0.06 (3.1) & 0.93$\pm$0.18 (6.3) &  2.15$\pm$0.36 (11.4) &  2.22$\pm$0.39 (11.1) &  1.81$\pm$0.36 (9.3) & 0.31$\pm$0.15 (2.1) \\
LBG-$M$4 & 10.50-10.75 & 10.59$\pm$0.27 & 10.61$\pm$0.08 & -0.82$\pm$0.52 & 2.95$\pm$0.26 &  492 & 0.38$\pm$0.12 (3.6) & 1.10$\pm$0.28 (4.7) &  3.01$\pm$0.56 (10.3) &  3.27$\pm$0.60 (10.1) &  2.67$\pm$0.75 (8.6) & 0.41$\pm$0.21 (2.0) \\
LBG-$M$5 & 10.75-11.00 & 10.57$\pm$0.29 & 10.86$\pm$0.08 & -0.80$\pm$0.53 & 2.93$\pm$0.26 &  213 & 0.47$\pm$0.15 (3.1) & <1.39         &  4.94$\pm$0.85 (11.2) &  5.56$\pm$0.95 (11.4) &  5.16$\pm$0.88 (10.0) & 1.19$\pm$0.42 (5.0) \\
LBG-$M$6 & 11.00-11.25 & 10.54$\pm$0.26 & 11.10$\pm$0.08 & -0.86$\pm$0.64 & 2.97$\pm$0.26 &   59 & 0.94$\pm$0.15 (3.4) & 3.33$\pm$0.87 (5.2) & 10.17$\pm$2.21 (12.1) & 11.51$\pm$2.41 (11.6) &  8.73$\pm$2.06 (9.8) & 2.05$\pm$0.92 (4.9) \\
\hline
\end{tabular}}
\tablefoot{The SNR of each stacked LBG in different bands and bins are presented in the table within brackets to the right of the fluxes and errors.}

 \label{TabLFs}
\end{table*}

Our selection has been carried out as a function of $L_{\mathrm{FUV}}$ after applying the colour selection criteria. For the stacking as a function of $L_{\mathrm{FUV}}$, we find statistical detections for most of the stacked bands and bins by making use of the total sample. For the stacking as a function of $\beta_{\mathrm{UV}}$ and $M_{*}$, we obtain no detection for the bins with $\beta_{\mathrm{UV}}$< -1.9 and log($M_{*}$[$L_{\odot}$])< 9.75. We would need a larger number of stacked LBGs in each bin to reduce the contribution of the background and obtain a statistical detection

The AzTEC observations cover around three times less area than the entire COSMOS field. We stack only sources in the covered region to compute the AzTEC (1.1mm) mean flux densities. We obtain a significant detection for the bins with, log($M_{*}$[L$_{\odot}$]) > 10 in the stacking as a function of stellar mass. As we stressed in Sect. \ref{lbgselection}, the source selection criteria are exactly the same inside and outside the covered area. We confirm that this does not introduce any bias in stacking analysis, by also stacking the AzTEC sub-sample in the PACS and SPIRE bands. The result of this stacking shows that the IR luminosity giving by the SED-fitting (Sect. \ref{ir_param}) are within the uncertainties in both cases.

\section{RESULTS AND DISCUSSION}
\textbf{
\subsection{IR luminosity}\label{ir_param}}

  \begin{figure*}
\begin{minipage}{.333\textwidth}
\includegraphics[width=\textwidth]{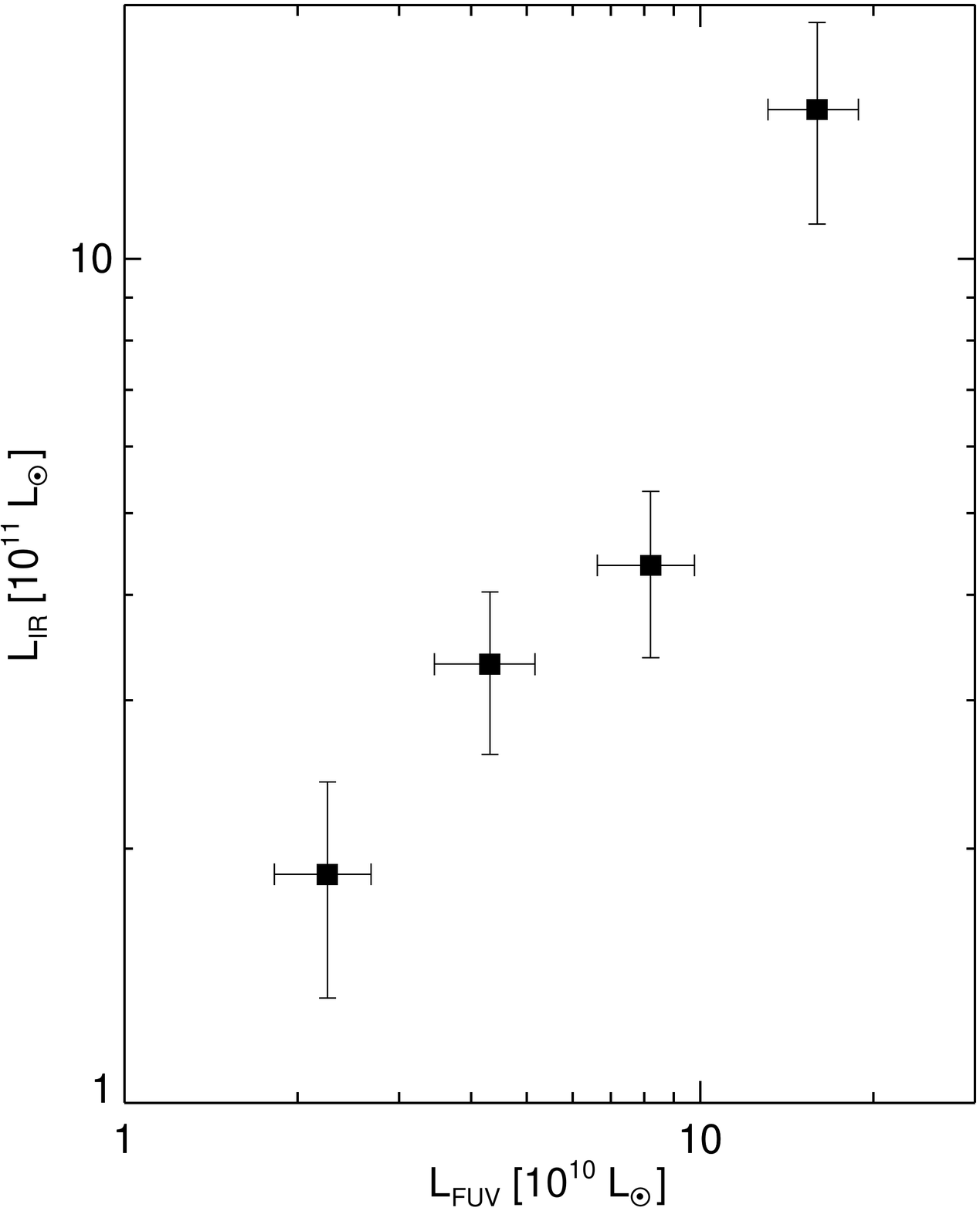}
\end{minipage}
\begin{minipage}{.333\textwidth}
\includegraphics[width=\textwidth]{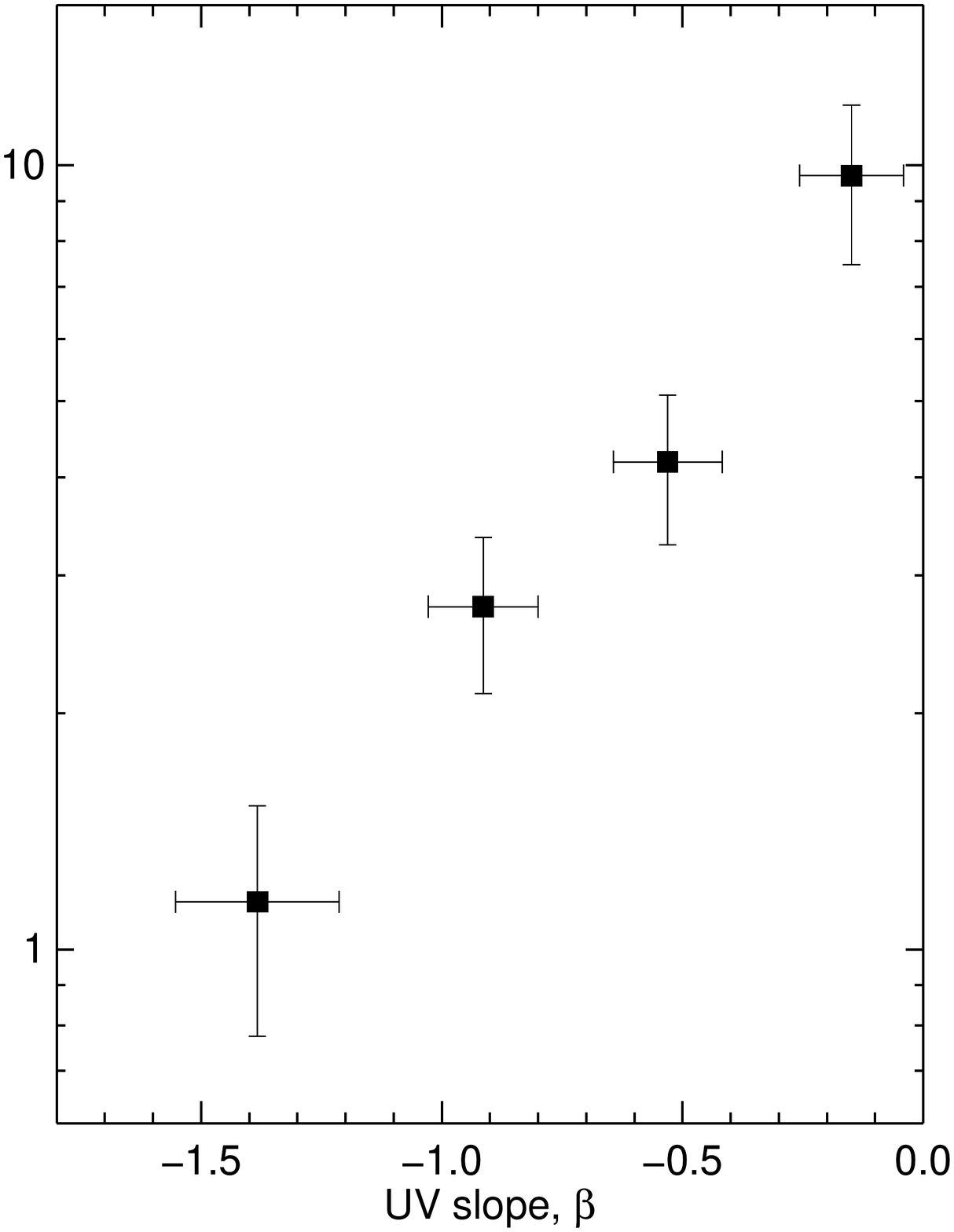}
\end{minipage}
\begin{minipage}{.333\textwidth}
\includegraphics[width=\textwidth]{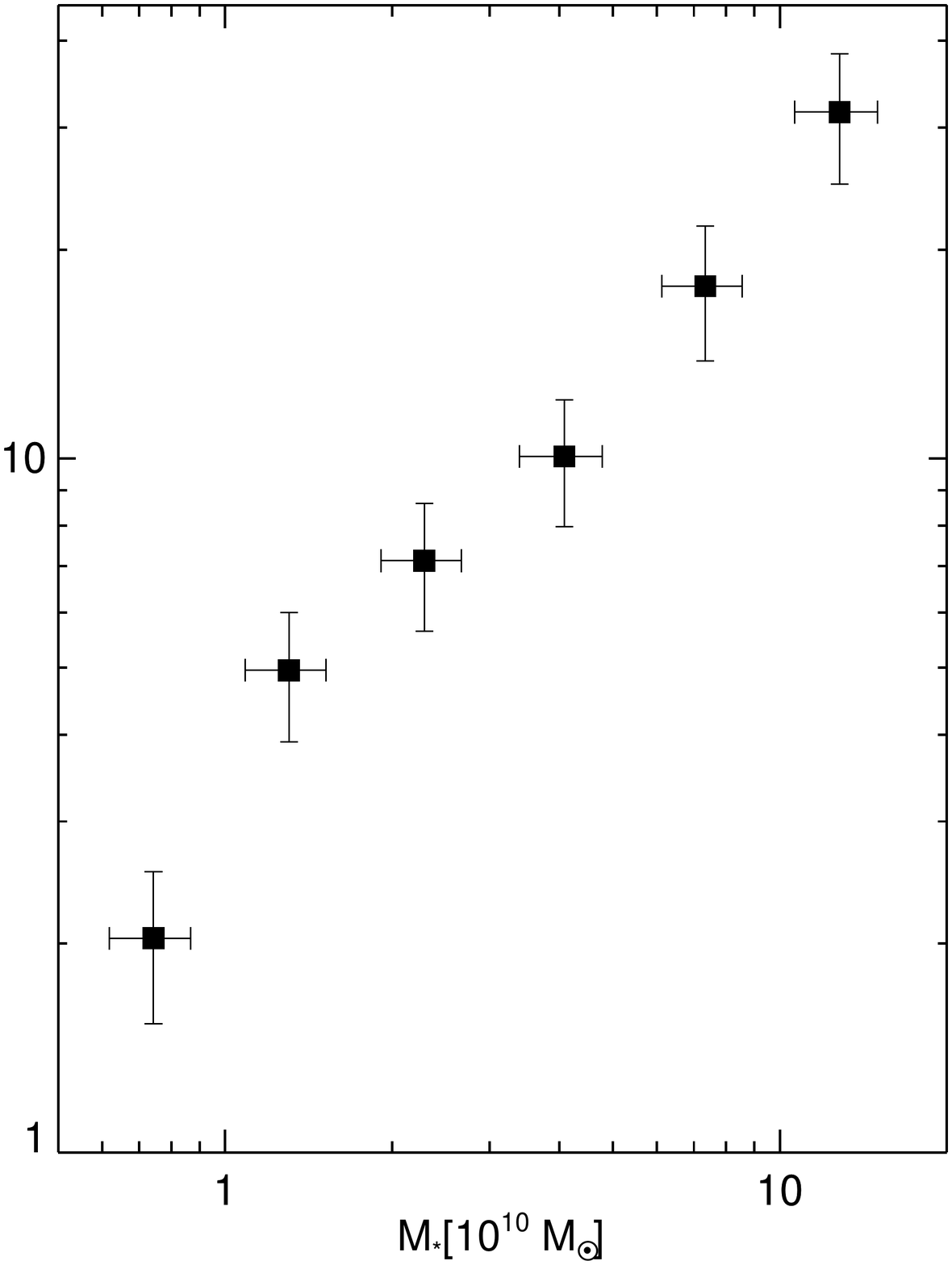}
\end{minipage}
      \caption{The average IR luminosity obtained for our stacked LBGs as a function of the $L_{\mathrm{FUV}}$ (left), $\beta_{\mathrm{UV}}$ (middle) and $M_{*}$ (right). }
         \label{L_IR}
   \end{figure*}
   
\begin{figure}[h]
   \centering
   \includegraphics[width=\hsize]{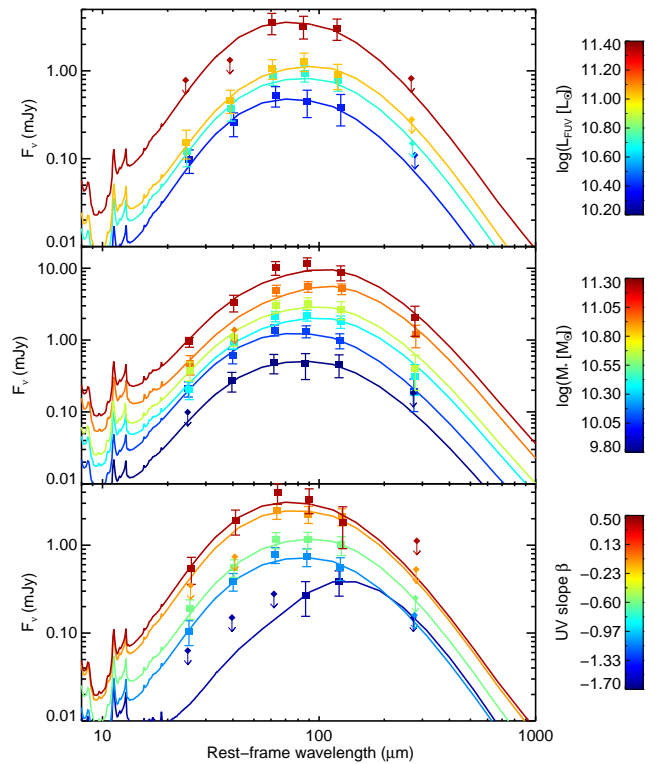}
      \caption{For the first time, we are able to obtain well-sampled stacked SED of the LBGs for each bin in $L_{\mathrm{FUV}}$ (top), $M_{*}$ (middle) and the $\beta_{\mathrm{UV}}$ (bottom). Data points are the average flux densities calculated from the stacking procedure in each band. The 3$\sigma$ upper limits (from bootstrap resampling) are shown using an arrow. We overplot the best-fit SED from the \cite{Dale2014} templates obtained using CIGALE.}
         \label{irfit}
   \end{figure}

We use the code CIGALE to fit the average stacked SED obtained for our LBGs to calculate the IR luminosity, which is estimated by integrating over the range $8 < \lambda < 1000~\mu$m the best-fit \cite{Dale2014} templates. X-ray AGN are removed from our LBG sample, since the mid-infrared emission of these objects could be strongly affected. X-ray obscured AGN might still be present in the sample. In this work, we do not have enough mid-IR data to constrain the AGN contribution for our stacked LBGs. However, \cite{Bethermin2014} have suggested that the possible presence of AGN emission appears to have limited impact on the stacking analysis of star-forming galaxies. For this reason, we use the \cite{Dale2014} templates with an AGN contribution fixed to 0\% and $\alpha_{\mathrm Dale}$ as a free parameter to compute the IR luminosity. Where $\alpha$ is the exponent of the power law of the dust mass distribution with the radiation field intensity $dU/dM \propto U^{-\alpha}$. Varying $\alpha$ provides a wide range of dust temperatures. The uncertainties are estimated by running a Monte Carlo simulations where each stacked LBGs is fit 3000 times. For each realization both the flux in each band and the mean redshift of each stacked LBGs are randomly perturbed from the actual values by drawing from a Gaussian distribution. The width is given by the flux uncertainty in the specific band in the first case and by the standard deviation of the redshift distribution calculated using the photo-$z$ for the objects in each bin in the second case. We take the standard deviation of the IR luminosities and $\alpha_{\mathrm Dale}$ obtained in each iteration as the error. Fig. \ref{irfit} shows IR SEDs and best models for our LBGs stacked as a function of their $L_{\mathrm{FUV}}$, their $\beta_{\mathrm{UV}}$ and their $M_{*}$ values.

The average IR luminosities obtained for our stacked LBGs cover the range between 3 $\times$ 10$^{10}$ to 3.23 $\times$ 10$^{12}$ $L_{\odot}$ (see Table \ref{t2}) and thus have luminosities in the range of Luminous Infrared Galaxies (LIRGs, 10$^{11}$ < $L_{\mathrm{IR}}$/$L_{\odot}$ < 10$^{12}$). However, there are some stacked LBGs with average IR emission in the Ultra Luminous Infrared Galaxy (ULIRG, 10$^{12}$ < $L_{\mathrm{IR}}$/$L_{\odot}$ < 10$^{13}$) range. Fig.~\ref{L_IR} shows the results of the average IR luminosities as a function of the $L_{\mathrm{FUV}}$, $\beta_{\mathrm{UV}}$ and $M_{*}$ values. The three relations follow a power-law where the IR luminosity increases as a function of the three parameters. 

%\textbf{\cite{Gruppioni2013} computed the IR LF at $z = 3$ using objects with typical luminosities in the ULIRG regime detected, in PACS and SPIRE images. The far-infrared, sub-mm and mm detected sources (SMGs using SCUBA by \citealt{Chapman05}, Herschel-SPIRE selected galaxies by \citealt{Casey2012}, LBGs detected in PACS by \citealt{Oteo2013} and aLESS sources detected with ALMA telescope by \citealt{Swinbank2014}) have IR luminosities larger than 1.8 $\times$ 10$^{12}$ L$_{\odot}$ in the redshift range $2.5 < z < 3.5$. Our sample instead tends to cover the lower luminosity regime in that observed LF. The bright end is typically well determined and known while the faint end of high-$z$ galaxies is quite uncertain, therefore the stacking analysis is important, because it allows us to characterize the faint LBGs population in a statistical way. }

The shape of the average IR SEDs and the best-fit models (Fig. \ref{irfit}) suggest that there is an evolution of the dust temperature as a function of $L_{\mathrm{FUV}}$, $\beta_{\mathrm{UV}}$ and $M_{*}$. Since $\alpha_{Dale}$ is inversely proportional to the dust temperature \citep{Dale2002, Chapman2003}, the value of $\alpha_{\mathrm dale}$ obtained reflect an evolution of the dust temperature (see Table \ref{t2}). On the one hand, for the stacking as a function of $L_{\mathrm{FUV}}$ and $M_{*}$, $\alpha_{Dale}$ increases, therefore, the average dust temperature for our LBGs decreases with these parameters. On the other hand, the average dust temperature shows an increasing trend with $\beta_{\mathrm{UV}}$.

\textbf{
\subsection{Dust attenuation}\label{att}}

It has been shown (e.g. \citealt{BuatXu1996}; \citealt{Gordon2000}) that the IR-to-UV luminosity ratio ($L_{\mathrm{IR}}/L_{\mathrm{FUV}}~\equiv$ IRX) is a robust tracer of the dust attenuation in star-forming galaxies. We define the FUV dust attenuation, A$_{FUV}$, as a function of the IRX according to the prescription by \cite{Meurer1999}:
\begin{equation}\label{eq_att2}
A_{FUV} = 2.5~\log  \left( \frac{BC_{dust}}{BC_{FUV,*}} ~IRX +1\right)
\end{equation}

where $A_{\mathrm{FUV}}$ is the dust attenuation in the FUV, $BC_{\mathrm{FUV},*}$ is the bolometric correction to the total light emitted by stars ($BC_{\mathrm{FUV},*}$ = 1.68, M99), $BC_{\mathrm dust}$ is the bolometric correction to the total light emitted by dust ($BC_{\mathrm dust}$=1). Despite the exact value of the bolometric correction for our specific sample of galaxies, we have chosen them to compare our best fit $A_{\mathrm{FUV}}$ to other works that use similar notation.
\\
\subsubsection{IRX-$\beta_{\mathrm{UV}}$ relation}\label{IRX_beta_relation}

The $\beta_{\mathrm{UV}}$ slope has been found to be a good tracer of the UV dust attenuation (e.g. \citealt{Meurer1999}).  It provides an estimate of the dust attenuation from the rest-frame UV without requiring far-IR data or spectral lines diagnostics. Calibrations have been derived from spectro-photometric samples of starburst galaxies at low redshift \citep{Meurer1999}. Further works \citep{Overzier2011,Takeuchi2012} have provided new estimates of the original M99 relation using \textit{Galaxy Evolution Exploter} (GALEX) data to compute the UV luminosity and solve the problem of the small aperture/field of view from the \textit{International Utraviolet Explorer} (IUE).  \cite{Takeuchi2012} also included a new $L_{\mathrm{IR}}$ estimation using AKARI data. They found lower IRX and redder $\beta_{\mathrm{UV}}$ values, for the same sample, than \cite{Meurer1999}. A more recent work by  \cite{Casey2015} proposed a new IRX-$\beta_{\mathrm{UV}}$ relation based on 1236 nearby galaxies observed by GALEX, for which they derive a much redder relation than the one presented by \citealt{Meurer1999} and roughly consistent with the aperture corrected M99 measured by \cite{Takeuchi2012}. Our intention is to investigate if the LBG population at $z\sim3$ follow, on average, the same IRX-$\beta_{\mathrm{UV}}$ relation by stacking them at Far-IR. 

%The main goal of \cite{Meurer1999} was to derive a relation using local starburst galaxies to measure the absorption-corrected $L_{\mathrm{FUV}}$ for LBGs at z$\sim$3, as previous work had shown that local starbursts and high-$z$LBGs have similar UV surface brightnesses \citep{Meurer1997}. Subsequent works also found that local ``supercompact'' UV-luminous galaxies match LBGs in terms of size, SFR, surface brightness, mass, metallicity, kinematics, dust, and UV-optical colour \citep{Overzier08}. Our intention is to investigate if the LBG population at $z\sim3$ follow, on average, the same IRX-$\beta_{\mathrm{UV}}$ relation by stacking them at Far-IR. 

The M99 relation has been widely used to estimate dust-corrected SFRs at high redshift as SFR estimates based on IR, X-ray, or radio data are only available for the brightest objects \citep{Bouwens2009}. However, galaxies of different types locate differently in the IRX-$\beta_{\mathrm{UV}}$ plane. Young, metal-poor galaxies like the SMC and LMC are redder and less dusty lain bellow the M99 relation. Dusty star-forming galaxies (DSFG) related with the IR-bright galaxy population lie instead above the M99 relation \citep{Casey2015, Oteo2013}. The dust geometry and star formation properties are thus seen to play a key role in the dispersion of the IRX-$\beta_{\mathrm{UV}}$. 

Fig.~\ref{irx_beta} shows the IRX-$\beta_{\mathrm{UV}}$ relation for our stacked LBGs as a function of $\beta_{\mathrm{UV}}$ (x-axis) and dust attenuation (right y-axis). The calibration from local starburst galaxies (M99), the aperture photometry corrected the M99 by {\citeauthor{Overzier2011} (2011, O11; hereafter we use the IRX$_{M99,total}$ derived in this paper), \citeauthor{Takeuchi2012} (2012, T12), and the new local calibration by \cite{Casey2015}, are also shown. In addition, we plot the LBGs detected in PACS at $z\sim3$ \citep{Oteo2013}, and the results of UV-selected galaxies at $z\sim1.5$ by \citealt{Heinis2013}). Our data points are, within the errors, in excellent agreement with the correction of the M99 relation by \citep{Takeuchi2012}. The comparison of our stacked LBGs with the stacking results of a large sample (42,184) of UV-selected galaxies at $z\sim1.5$ by \cite{Heinis2013} confirms that we do not see large evolution in the IRX-$\beta_{\mathrm{UV}}$ relation for average populations of galaxies selected from UV colours (LBGs and UV-selected), and from redshift 1.5 to 3. However, the IRX-$\beta_{\mathrm{UV}}$ relation derived from local galaxies by \cite{Casey2015} presents different slope, but most of our data points present an agreement, within the uncertainty.% They cover a more heterogeneous population of galaxies which also includes objects with much lower SFRs than typical LBGs. These galaxies could contain a lower number of OB stars resulting in a redder UV spectrum (the higher metallicity and age of the stellar population can also contribute).    

   \begin{figure}[h]
   \centering
   \includegraphics[width=\hsize]{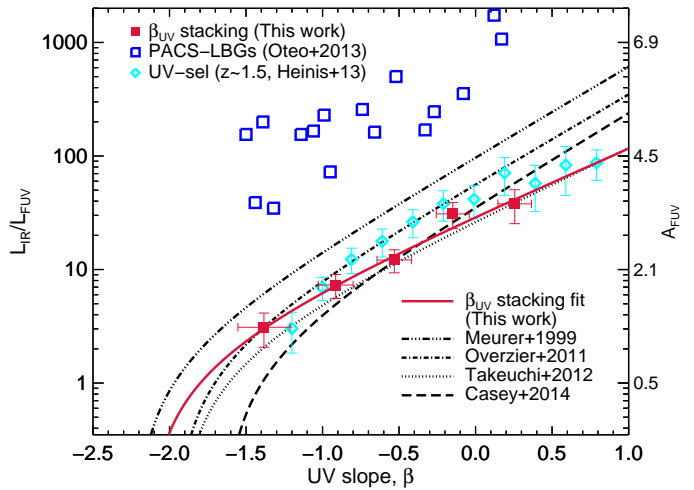}
      \caption{IRX-$\beta_{\mathrm{UV}}$ diagram. The right-hand axis shows the equivalent attenuation in the FUV band, in magnitudes, computed using Eq. \ref{eq_att2}. Our data points  and the best-fit to Eq. \ref{eq_att2} are shown in red squares and red solid line, respectively. The gray circles are the results of the stacking as a function of $M_{*}$, the size of the uncertainty in the x-axis is show in the legend. Lines show various IRX-$\beta_{\mathrm{UV}}$ relations: the local calibration of M99 (dot-dot-dot-dashed line); the aperture correction of the M99 relation by T12 (doted line) and O11 (solid line), and the new local calibration by \citeauthor{Casey2015} (dashed line, 2015). We show measurements at $z\sim3$ from direct LBGs detection in PACS by \citeauthor{Oteo2013} (Blue open square, 2013), and the results of UV-selected galaxies at z$\sim$1.5 \citep{Heinis2013}. The result from this work are in good agreement with \cite{Takeuchi2012}. Oteo's detected LBGs seem to be extracted from a biased IR-bright LBGs, not representative of the average population. It is interesting to notice that $A_{FUV}$ continuously increases with increasing stellar masses.}          
         \label{irx_beta}
   \end{figure} 

The dust attenuation in Eq. \ref{eq_att2} can be written as a function of $\beta_{\mathrm{UV}}$: $A_{FUV}$ = $C_{0}$+$C_{1}$ $\beta_{\mathrm{UV}}$ \citep{Meurer1999,Overzier2011}. We obtain as best-fitting parameters for LBGs at $z\sim3$, $C_{0}$=3.15$\pm$0.12 and $C_{1}$=1.47$\pm$0.14. This equation implies that the UV slope of the dust-free objects is $\beta_{\mathrm{dust-free}} = -2.2 \pm 0.3$ being in agreement, within the errors, what is expected from stellar population models and a constant SFH mode \citep{Leitherer1995}. Note that the $\beta_{\mathrm{UV}}$ range study in this work does not allow to constrain the bluer part of the IRX-$\beta_{\mathrm{UV}}$ relation, which is really important to determine the UV slope of a dust-free population.

Previous works have found the local starburst relation (M99) to hold for LBGs at various redshifts. \cite{Magdis2010c} found that the dust corrected UV-SFR derived from M99 relation presents a good match with the far-IR and radio SFR estimators by stacking an IRAC spectroscopically confirmed LBG sample at $z\sim3$ in MIPS (24~$\mu$m), AzTEC (1.1~mm) and radio (1.4~GHz). This stacking analysis of spectroscopically confirmed $z\sim2$ LBGs in Far-IR \citep{Reddy2012}, $z\sim4$ LBGs at radio continuum (1.4GHz, \citealt{To2014}),  and direct detection of LBGs in PACS at lower redshift ($z\sim1$, \cite{Oteo2013a}) also lie on the M99 relation. However, if we compare the IRX-$\beta_{\mathrm{UV}}$ relation obtained here with the original M99 relation, the previous work present bluer colors and/or higher dust attenuation than our mean LBGs population. \cite{Buat2015} showed that the selection sample has an influence in the mean dust attenuation. They obtained differences up to 2 mag between the UV-selected and IR-selected sample. In the particular case of \cite{Magdis2010c}, their sample is IRAC-selected LBGs which is probably the origin of the higher dust attenuation. In Sect. \ref{IRX_beta_mass}, we will investigate with more details the effect in the dust attenuation due to the definition of our sample.

The  LBGs detected in PACS at $z\sim3$ \citep{Oteo2013} are found to be outliers of the IRX-$\beta_{\mathrm{UV}}$ relation. These galaxies are mainly ULIRG and Hyper-luminous infrared galaxies (HLIRG, 10$^{13}$ < $L_{\mathrm{IR}}$/$L_{\odot}$ < 10$^{14}$), similar to the DSFG and thus not representative of the average LBG population. \cite{Casey2015} showed that there is a deviation of this relation towards bluer colors for galaxies with $L_{\mathrm{IR}}$> 10$^{11-11.5}$ due to the presence of recent and fast episodes of star-formation that produce more prominent population of young O-B stars (contributing to the rest-frame far-UV emission) than galaxies of more modest SFRs. If we analyze the stacking results as a function of stellar mass in the IRX-$\beta_{\mathrm{UV}}$ plane, we find that they present bluer colours than our IRX-$\beta_{\mathrm{UV}}$ relation for the bins with $\log (M_{*}~[L_{\sun}]) > 10.25$ and $L_{\mathrm{IR}}>7 \times 10^{11}~L_{\sun}$, showing the same behavior than the DSFG by \cite{Casey2015}. The stellar mass content in the galaxy could be the main driver of the dispersion of the IRX-$\beta_{\mathrm{UV}}$ plane to bluer colours, or simply could be a consequence of the fact that more massive LBGs have associated larger infrared luminosity at high redshift.

\subsubsection{ IRX-$M_{*}$ relation}\label{IRX_mass_relation}

The stellar mass has been shown to correlate with the dust attenuation in LBGs \citep{Reddy2010}, UV-selected galaxies \citep{Buat2012, Heinis2014} and a mass-complete sample of galaxies \citep{Pannella2015}. The IRX-$M_{*}$ relation presents no significant evolution with redshift, \cite{Pannella2015} found less than 0.3 magnitudes of difference in the $A_{FUV}$ from redshift $\sim0.7$ to $\sim3.3$, and \cite{Heinis2014} also showed consistent results from $z\sim1.5$ to 4. In Sect. \ref{IRX_beta_relation}, we have shown that $\beta_{\mathrm{UV}}$ presents a correlation with the dust attenuation. In fact the average $\beta_{\mathrm{UV}}$ for a population of galaxies and $M_{*}$ clearly correlate at all redshifts (also for our LBGs sample, see Table \ref{t1}), but while the attenuation is fairly constant, or slightly increasing with redshift, $\beta_{\mathrm{UV}}$ becomes systematically bluer. The fact that high mass galaxies at high redshift have a similar dust attenuation but a bluer $\beta_{\mathrm{UV}}$ compared to similar mass galaxies at lower redshift, has important implications for UV-derived SFRs in the high redshift Universe \citep{Pannella2015}. This leads to an inconsistency between dust attenuation measurements obtained using $\beta_{\mathrm{UV}}$ or $M_{*}$. Fig.~\ref{irx_mass} shows our measurements for the IRX as a function of $M_{*}$. We also plot the relation from UV-selected galaxies by \cite{Heinis2014}, and the one from a complete sample of star-forming galaxies by \cite{Pannella2015} and the LBGs detected in PACS at $z\sim3$ \citep{Oteo2013}.  

   \begin{figure}[h]
   \centering
   \includegraphics[width=\hsize]{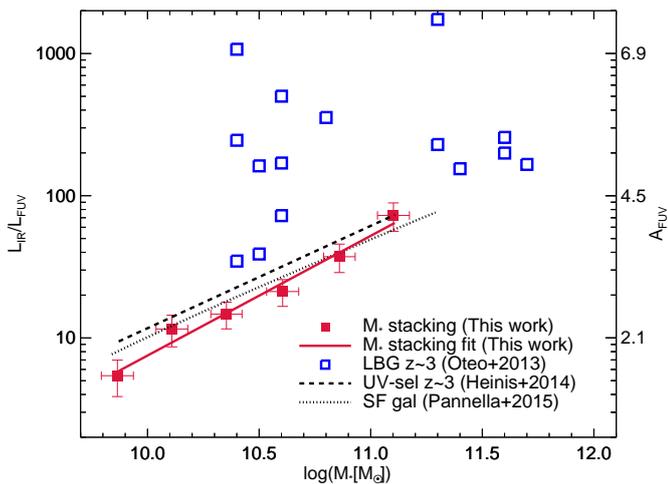}
      \caption{IRX versus $M_{*}$. The right-hand axis shows the equivalent attenuation in the FUV band, in magnitudes, obtained using Eq.~\ref{eq_mass}. Our data points  and the best-fit to Eq.~\ref{eq_att2} are shown in red squares and red solid line, respectively. We show previous measurements at $z\sim3$ from the stacking of UV selected galaxies by \citeauthor{Heinis2014} (dashed line, 2014), star-forming galaxies valid for redshift range 0.5-4 by \citeauthor{Pannella2015} (dotted line, 2015) and direct LBGs detections in PACS by \citeauthor{Oteo2013} (blue open square, 2013).}
         \label{irx_mass}
   \end{figure}

We assume here the following relation between the IRX and stellar mass:
\begin{equation}\label{eq_mass}
\log (IRX) = \alpha \log \left( \dfrac{M_{*}}{10^{10.35}} \right) + IRX_{0}
\end{equation}

we obtain as best-fitting parameters, $\alpha$ =0.84$\pm$0.11 and IRX$_{0}$ =1.17$\pm$0.05 .

We compare our results with previous published laws. We find that our IRX-$M_{*}$ relation presents a steeper slope than \cite{Heinis2014} and \cite{Pannella2015} ones. However, \cite{Pannella2015}  relation presents a good agreement with most of our data points in the range from $10<\log(M_{*}[M\sun])<11.25$. The difference could stem from the incompleteness in the low stellar mass bin that could present some variation in the dust attenuation and SFR with respect to a complete sample of galaxies. 

In this case, the IRX-$M_{*}$ relation tends to be aligned with the most massive ($\log(M_{*}[M\sun])>11)$ LBGs detected in PACS by \cite{Oteo2013}. However, the lower stellar mass objects ($\log(M_{*}[M\sun])<11$) are located above the IRX-$M_{*}$ relation showing large dust attenuation and/or the possible dispersion in the IRX-$M_{*}$ plane due to the different nature of the objects.

As already mentioned in Sect. \ref{IRX_beta_relation} and in the first part of this Sect., the results obtained as a function of $\beta_{\mathrm{UV}}$ and $M_{*}$ present different behaviors in the IRX-$\beta_{\mathrm{UV}}$ and IRX-$M_{*}$ planes. The more massive LBGs in the stacking as a function of $M_{*}$ suggest larger dust attenuation and bluer $\beta_{\mathrm{UV}}$ than the stacking as a function of $\beta_{\mathrm{UV}}$. %This can be because of the much lower ($9.65 <\log(M_{*}[M_{\sun}])<9.92$) mean of $M_{*}$ for each bin of the stacking as a function of $\beta_{\mathrm{UV}}$. Here, we investigate a different population of LBGs. It might also be because the most massive LBGs present higher IR luminosity and bluer $\beta_{\mathrm{UV}}$ as argued by \cite{Casey2015}. We stress that the definition of the sample can strongly impact the dust attenuation \citep{Buat2015}. It is therefore crucial to make use of the best IRX relation depending on the studied sample.

\subsubsection{Comparison between IRX-$M_{*}$ and IRX-$\beta_{\mathrm{UV}}$ relations}\label{IRX_beta_mass}

  \begin{figure*}
\begin{minipage}{.4999\textwidth}
\includegraphics[width=\textwidth]{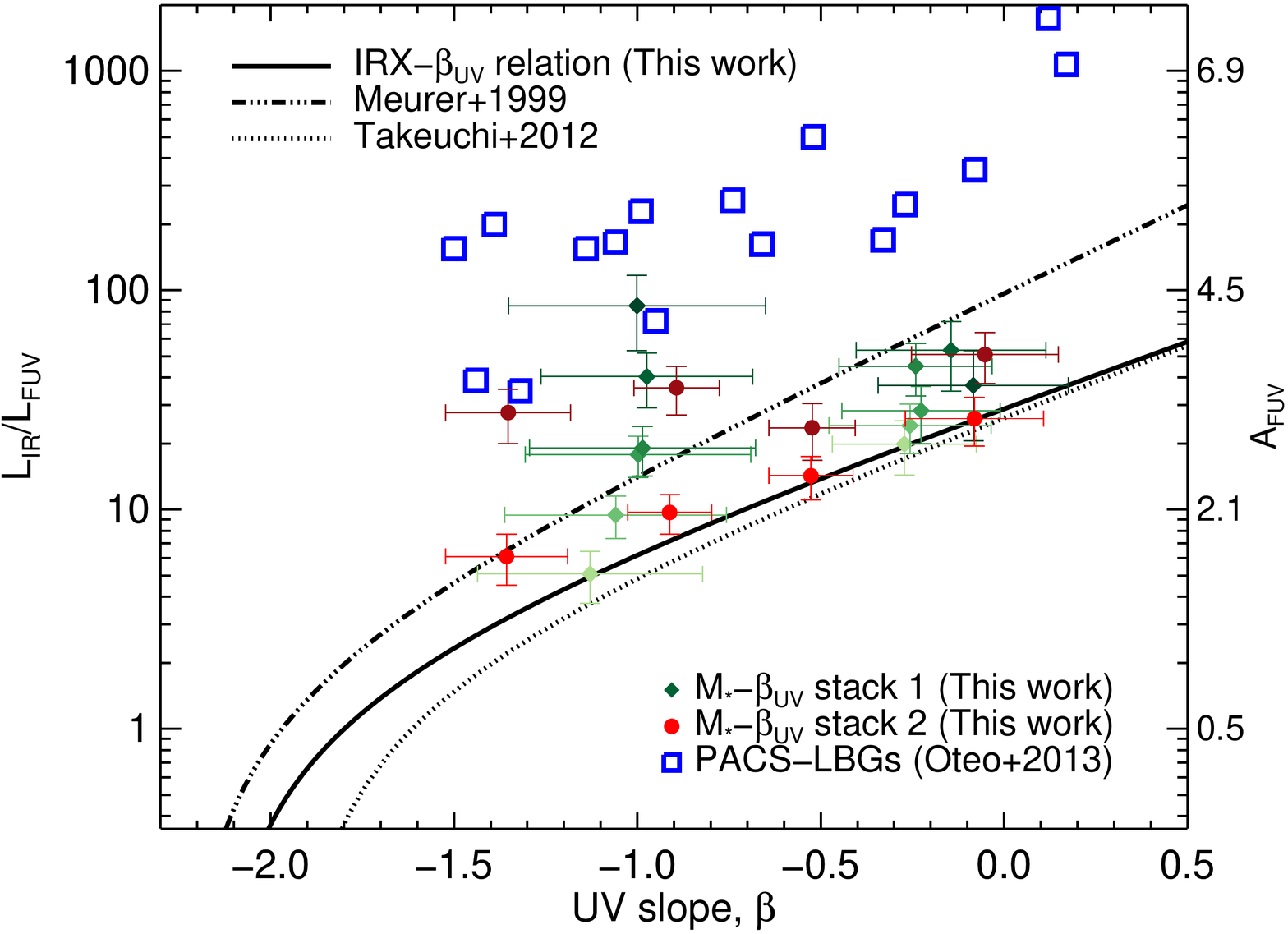}

\end{minipage}
\begin{minipage}{.4999\textwidth}
\includegraphics[width=\textwidth]{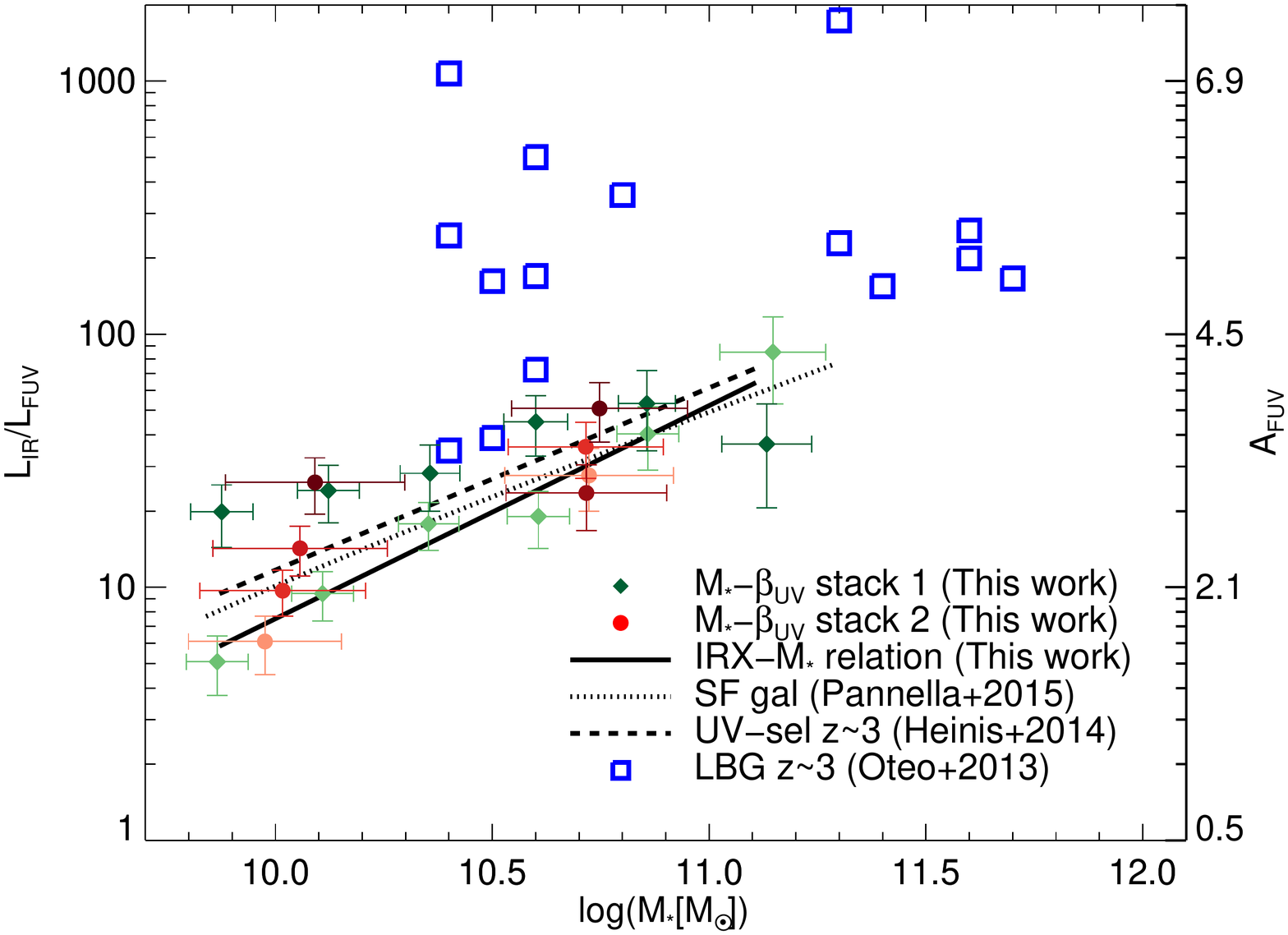}

\end{minipage}
      \caption{it shows the stack 1 (green) and stack 2 (red) results in the IRX-$\beta_{\mathrm{UV}}$ and IRX-$M_{*}$ plane. Left panel: IRX-$\beta_{\mathrm{UV}}$ diagram, for reference see Fig. \ref{irx_beta}. The stack 1 results are showed in diamonds, the tonalities of green represent the increase of the $M_{*}$ from 9.75 (light green) to 11.50 (dark green). The filled red circles are the results from the stack 2, the tonalities of red represent the two different bins in $M_{*}$, $9.75 < \log(M_{*}[M_{\sun}]) < 10.5$ (light red) and $10.5 < \log(M_{*}[M_{\sun}]) < 11.5$ (dark red). Right panel: IRX-$M_{*}$ diagram, for reference see Fig. \ref{irx_mass}. The stack 1 results are showed in diamonds, the tonalities of green represent the two different bins in $\beta_{\mathrm{UV}}$,  $-1.7 < \beta_{\mathrm{UV}} < -0.5$ (light green) and $-0.5 < \beta_{\mathrm{UV}} < 0.5$ (dark green). The filled red circles are the results from the stack 2, the tonalities of red represent the increase of the $\beta_{\mathrm{UV}}$ from -1.7 (light red) to 0.5 (dark red).}
         \label{dispersion}
   \end{figure*}

In Sect. \ref{IRX_beta_relation} and \ref{IRX_mass_relation}, we presented the mean IRX-$M_{*}$ and IRX-$\beta_{\mathrm{UV}}$ relations for our LBG population. These previous stacking analyses do not provide any information to quantify the possible dispersion in the IRX relations. We also notice that both relations are not fully consistent, the more massive LBGs do not follow the IRX-$\beta_{\mathrm{UV}}$ relation presenting larger dust attenuation and bluer colours. We investigate here the dispersion and nature of the differences between the IRX-$\beta_{\mathrm{UV}}$ and IRX-$M_{*}$ relations. We perform two new stacking analyses by building in different way sub-samples in the  ($M_{*}$, $\beta_{\mathrm{UV}}$) plane. In one hand, we split our sample in six bins of  $M_{*}$ and each of these bins is further separated in two bins of $\beta_{\mathrm{UV}}$ (stack 1)\footnote{Stack 1: the size of the bin as a function of $M_{*}$ are 0.25 dex from 9.75 to 11.00, and we include a last bin form 11.0 to 11.50, where the $M_{*}$ is defined as $\log(M_{*}[M_{\sun}])$. The two bins in $\beta_{\mathrm{UV}}$ are: $-1.7 < \beta_{\mathrm{UV}} < -0.5$ and $-0.5 < \beta_{\mathrm{UV}} < 0.5$.}. On the other hand, we split our sample in 4 bins of  $\beta_{\mathrm{UV}}$ and each of these bins is  further separated in two bins of $M_{*}$ (stack 2)\footnote{Stack 2: The size of the bin as a function of $\beta_{\mathrm{UV}}$ are [0.6, 0.4, 0.4, 0.8] from -1.7 to 0.5. The two bins in $M_{*}$ are: $9.75 < \log(M_{*}[M_{\sun}]) < 10.5$ and $10.5 < \log(M_{*}[M_{\sun}]) < 11.5$.}. In these new stacking analyses, we limit the sample in the intervals as a function of $M_{*}$ ($9.75 <\log(M_{*}[M_{\sun}])<11.5$) and $\beta_{\mathrm{UV}}$ ($-1.7 < \beta_{\mathrm{UV}} < 0.5$) obtaining around 9,000 LBGs. We stack following the same procedure than Sect. \ref{stackinganalisis}. We define this binning with the objective that all the stacked LBGs be detected in SPIRE, because these bands are the main bands to compute the IR luminosity. We also stack in PACS and find a detection in part of them, but we did not use AzTEC in this analysis because we reach no detection. Fig. \ref{dispersion} presents the results for the both stacks in the IRX-$\beta_{\mathrm{UV}}$ and IRX-$M_{*}$ planes.

In the IRX-$\beta_{\mathrm{UV}}$ plane, the stack 1 present a trend from the mean IRX-$\beta_{\mathrm{UV}}$ relation to higher IRX values and/or bluer colours. The lowest stellar mass bin has the same IRX than the mean IRX-$\beta_{\mathrm{UV}}$ relation, and the highest stellar mass bin tends to lie near the LBGs detected in PACS by \cite{Oteo2013}. This trend is well defined in the bluer bins, but the redder bins present a large dispersion due to the low number of objects. The stack 2 shows the same behavior than the stack 1, when the stellar mass increases the IRX presents large values. These results suggest that the mean IRX-$\beta_{\mathrm{UV}}$ relation is not a well-defined relation valid for any objects by rather presents a large dispersion, which could be due to the differences in the stellar mass content in the galaxy for a given $\beta_{\mathrm{UV}}$. \cite{Casey2015} proposed that the origin of the effect might be sought in the IR luminosity, but we show here that the $M_{*}$ can also be the main driver. We should consider that the IR luminosity increases with $M_{*}$, therefore the IR luminosity and $M_{*}$ correlate for our LBGs at z$\sim3$ (see Fig. \ref{L_IR}). So, both the IR luminosity and the stellar mass have an influence in the dispersion of the IRX-$\beta_{\mathrm{UV}}$ diagram.

The mean  IRX-$\beta_{\mathrm{UV}}$ relation, obtained in Sect. \ref{IRX_beta_relation}, is dominated for low $M_{*}$ population ($9.70 <\log(M_{*}[M_{\sun}])<10.12$, see Table \ref{t1}). In the IRX-$\beta_{\mathrm{UV}}$ plane, the low stellar mass bin ($\log(M_{*}[M_{\sun}] \sim 10.05$) for the stack 2 presents a good correlation with the mean IRX-$\beta_{\mathrm{UV}}$ relation. However, the high stellar mass bin ($\log(M_{*}[M_{\sun}] \sim 10.75$) shows a flat behavior with large IRX values. This means that the blue and high stellar mass  LBGs lie above the mean IRX-$\beta_{\mathrm{UV}}$ relation, located in the same area than the LIRGs and ULIRGs DSFG from \cite{Casey2015}.

In the IRX-$M_{*}$ plane, for the stack 1, the bluer bins are in agreement with the previous results in the stacking as a function of $M_{*}$. The redder bins are systematically above the mean IRX-$M_{*}$ relation, the lower stellar mass bins show larger differences than the higher stellar mass bins which are in agreement with the mean IRX-$M_{*}$ relation. For the stack 2,  the low $M_{*}$ bins present a dust attenuation departure from the IRX-$M_{*}$ relation up to 1.3 mag higher. However, the high $M_{*}$ bins are scattered around the mean IRX-$M_{*}$ relation.

In the stack 1, the dust attenuation shows a departure by up to 2.8 mag above the mean IRX-$\beta_{\mathrm{UV}}$ relation, for the same $\beta_{\mathrm{UV}}$ bin when the $\log(M_{*}[M_{\sun}])$ increases from 9.75 to 11.5.  In the stack 2, the dust attenuation also shows a departure by up to 1.3 mag above the mean IRX-$M_{*}$ relation, for the same $M_{*}$ bin when $\beta_{\mathrm{UV}}$ increases from -1.7  to 0.5. Previous works also showed that IR-selected galaxies have a mean dust attenuation $\sim$2 magnitudes higher than the UV-selected galaxies at $z\sim0-2$ \citep{Buat2015}. We suggest that the criterion in the selection of the sample has a strong impact in the mean dust attenuation of the population. 

We would like to stress that the IRX-$M_{*}$ plane presents a lower dispersion than the IRX-$\beta_{\mathrm{UV}}$ for $M_{*}$ and $\beta_{\mathrm{UV}}$ intervals investigate here. However, we find that the objects with low stellar mass LBGs ($\log(M_{*}[M_{\sun}]) < 10.5$) and red $\beta_{\mathrm{UV}}$ ($\beta_{\mathrm{UV}} > -0.7$), $\sim$ 15\% of the total sample, present larger dust attenuation than the mean IRX-$M_{*}$, but they are in agreement with the mean IRX-$\beta_{\mathrm{UV}}$ relation. We suggest that we have to combine both, IRX-$\beta_{\mathrm{UV}}$ and IRX-$M_{*}$, relations to obtain the best estimation of the dust attenuation from the UV and NIR properties of the galaxies ($L_{\mathrm{FUV}}$, $\beta_{\mathrm{UV}}$, $M_{*}$). 

\subsubsection{IRX-$L_{\mathrm{UV}}$ relation}

We investigate how the stacking as a function of $L_{\mathrm{UV}}$ holds in the IRX-$L_{\mathrm{UV}}$ plane. Previous works show that the IRX remains more or less constant for different bins of $L_{\mathrm{UV}}$ in the average population of UV-selected galaxies. \cite{Xu2007} presented a sample of 600 UV-selected galaxies in GALEX with $z\sim0.6$ and the stacking analysis in SWIRE, they obtained a constant IRX ($\log(IRX)\sim0.8$) for the different bins in $L_{\mathrm{UV}}$. More recent works \citep{Heinis2013, Heinis2014} studied a large sample of  UV-selected galaxies at $z\sim$~1.5, 3 and 4 in the COSMOS field by stacking them in SPIRE images. They found that the IRX remaind more or less constant for the different bins in $L_{\mathrm{UV}}$ at  $z\sim1.5$, but they obtain a trend to higher dust attenuation for lower $L_{\mathrm{UV}}$ at high redshift. However, the IRX-$L_{\mathrm{UV}}$ plane presents a large dispersion for individual galaxies, \cite{Buat2015} showed that IR selected have a dust attenuation $\sim$2 magnitudes higher than that the UV-selected galaxies at $z\sim0-2$.

Fig.~\ref{irx_lum} shows the IRX-$L_{\mathrm{FUV}}$ relation for our stacking sampling as a function of $L_{\mathrm{FUV}}$. The IRX is found to be roughly constant over the range of $L_{\mathrm{FUV}}$ that we probe, with a mean of 7.9. This was already expected from the Fig. \ref{L_IR}, that show a lineal relation between $L_{\mathrm{FUV}}$ and $L_{\mathrm{IR}}$. Our results are in agreement with the stacking analysis of a sample of UV-selected galaxies at $z\sim3$ by \cite{Heinis2014}. They suggested that the IRX-$L_{\mathrm{FUV}}$ relation presents a trend, but we can not independently confirm it due to the fact that the IRX uncertainty are in agreement with constant behavior and we are missing the faintest $L_{\mathrm{FUV}}$ bin. \cite{Coppin2015} present a stacking analysis of the whole sample (around 4200) of $z\sim3$ LBGs in SCUBA2, SPIRE, 24$\mu$m and radio (1.4GHz) finding an average IRX around 8.  However, the spectroscopic sample of LBGs by \cite{Reddy2012} has lower IRX, 7.1 $\pm$ 1.1, but consistent within 1$\sigma$ error. We are also in agreement with the calculation of the dust attenuation of \cite{Burgarella2013}, who found an IRX equal to 7.59$\pm$6.99 and 5.54$\pm$5.59 for galaxies at redshifts $\sim$ 2.71 and 3.15, respectively.

These results suggest that for studies investigating the average population of LBGs can use a constant dust attenuation as a function of  $L_{\mathrm{FUV}}$ to correct the observed $L_{\mathrm{FUV}}$. For example, our mean dust attenuation could be used to derive the contribution of the LBGs at $z\sim3$ to the star formation rate density by correcting the observed UV luminosity function (e.g. \citealt{Madau2014}). However, there is a large dispersion on the IRX-$L_{\mathrm{FUV}}$ plane and, as commented before, the different sample selections present different dust attenuation \citep{Buat2015}. If we compare the stacking results as a function of the $\beta_{\mathrm{UV}}$ and $M_{*}$, where we are exploring the IRX-$L_{\mathrm{FUV}}$ plane along the IRX axis, they present a variation on the dust attenuation form 1.4 to 3.8 and 0.5 to 2.9 magnitudes, respectively. This give a good view of dispersion of the IRX-$L_{\mathrm{FUV}}$ plane, when we use different selections inside of the same sample of galaxies, however we are losing the information of the $L_{\mathrm{FUV}}$ axis. This dispersion is because we can find the same $L_{\mathrm{FUV}}$ for a galaxy with large IR emission and dust attenuation (red galaxy) than for a galaxy that have low IR emission and dust attenuation (blue galaxy). 

As we already discussed, our results suggest that the dust attenuation is roughly constant with $L_{\mathrm{FUV}}$ or equivalently with M$_{FUV}$. However, there is a debate on whether the $\beta_{\mathrm{UV}}$ evolves with the $L_{\mathrm{FUV}}$  at high redshift \citep{Bouwens2014,Bouwens2012,Bouwens2009, Finkelstein2012}. In one hand, \cite{Bouwens2009,Bouwens2014} show a trend between the $\beta_{\mathrm{UV}}$ as a function of the M$_{FUV}$ from redshift 2 to 8, they relate this to the metallicity and/or dust attenuation evolution. On the other hand, \cite{Finkelstein2012} found a flat distribution of the $\beta_{\mathrm{UV}}$ as a function of the $M_{\mathrm{FUV}}$, but a trend with the stellar mass. For our sample of LBGs at $z\sim3$, we show that the $\beta_{\mathrm{UV}}$ value roughly constant for each bin in the stacking as a function of the $L_{\mathrm{FUV}}$, consistent with the  \cite{Finkelstein2012} results. Therefore, we can conclude after seeing the differences between our results and these from \cite{Bouwens2009,Bouwens2014} in the $\beta_{\mathrm{UV}}$-$L_{\mathrm{FUV}}$ relation, that we cannot determine if the trend in the $\beta_{\mathrm{UV}}$-$L_{\mathrm{FUV}}$ relation is coming from evolution in dust attenuation and/or a bias in the sample selection.

   \begin{figure}[h]
   \centering
   \includegraphics[width=\hsize]{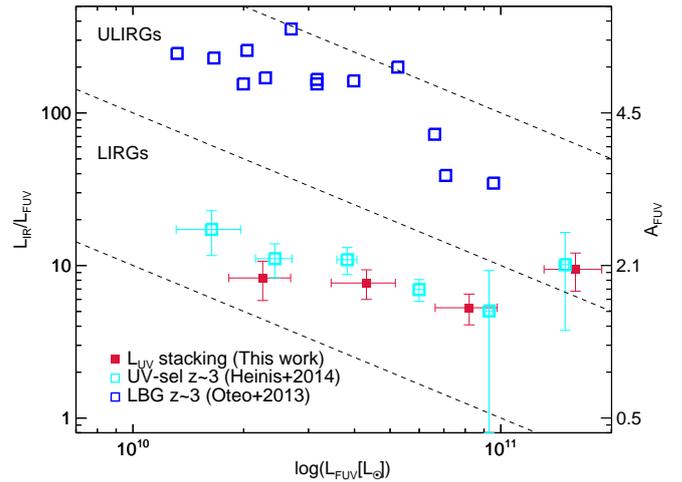}
      \caption{IRX versus $L_{\mathrm{FUV}}$.The right-hand axis shows the equivalent attenuation in the FUV band, in magnitudes, using Eq.~\ref{eq_att2}. Our data points are shown in red square. We also show previous measurements at $z\sim3$ from stacking of UV-selected galaxies in the COSMOS field by \citeauthor{Heinis2014} (cyan open squares, 2014) and LBGs detected in PACS by \citeauthor{Oteo2013} (blue open square, 2013). The dashed lines represent the region where LIRG and ULIRG lie.}
         \label{irx_lum}
   \end{figure} 

\textbf{
\subsection{Star-formation rate (SFR)}\label{sfr}} 

Since we have all the information to estimate reliable infrared luminosity for LBGs, we can combine the average estimates of $L_{\mathrm{IR}}$ and the observed uncorrected average UV luminosities to compute the total star-formation rates (SFR = SFR$_{\mathrm{FUV}}$ + SFR$_{\mathrm{IR}}$). We adopt the calibrations of \cite{Kennicutt1998}
\begin{equation}\label{eq_sfrIR}
SFR_{\mathrm{IR}}(M_{\odot}yr^{-1}) = 1.10 \times 10^{-10} L_{\mathrm{IR}}(L_{\odot})
\end{equation}
\begin{equation}\label{eq_sfruv}
SFR_{\mathrm{FUV}}(M_{\odot}yr^{-1}) = 1.82 \times 10^{-10} L_{\mathrm{FUV}}(L_{\odot})
\end{equation}
rescaled from \cite{Salpeter1955}  to a \cite{Chabrier2003} IMF.\\
\begin{table}

\caption{\label{t2}Physical parameters for the stacked LBGs}
\centering
\resizebox{0.47\textwidth}{!}{
 \renewcommand{\arraystretch}{1.5}
\begin{tabular}{cccccc}
\hline
ID & L$_{IR}$[10$^{11}$ L$_{\sun}$] &  SFR$_{tot}$ [M$_{\sun}$yr$^{-1}$] & IRX & A$_{FUV}$ & $\alpha_{Dale}$ \\
\hline
\multicolumn{6}{c}{\textbf{Stacking as a function of $L_{\mathrm{FUV}}$ (LBG-$L$)}}\\
\hline

LBG-L1 & 1.87$\pm$0.54 &  24.6$\pm$5.8  &  8.3$\pm$2.4  & 1.9$\pm$0.6 & 1.3$\pm$0.4\\

LBG-L2 & 3.31$\pm$0.72 &  44.3$\pm$7.9  &  7.7$\pm$1.7 & 1.9$\pm$0.4 & 1.5$\pm$0.2\\

LBG-L3 & 4.34$\pm$0.97 &  62.6$\pm$10.7 &  5.3$\pm$1.2 & 1.5$\pm$0.5 & 1.5$\pm$0.3\\

LBG-L4 & 15.04$\pm$4.04 & 194$\pm$45    &  9.4$\pm$2.6 & 2.1$\pm$0.6 & 1.4$\pm$0.5\\

\hline
\multicolumn{6}{c}{\textbf{Stacking as a function of $\beta_{\mathrm{UV}}$ (LBG-$\beta$)}}\\
\hline

LBG-$\beta$1 &  1.15$\pm$0.38  & 19.4$\pm$4.1  &  3.1$\pm$1.1   & 1.1$\pm$0.6 & 2.2$\pm$0.5\\
LBG-$\beta$2 &  2.73$\pm$0.62  & 36.9$\pm$6.8  &  7.3$\pm$1.8   & 1.8$\pm$0.5 & 1.4$\pm$0.3\\
LBG-$\beta$3 &  4.18$\pm$0.90  & 52.3$\pm$10.0 &  12.1$\pm$2.3  & 2.3$\pm$0.5 & 1.5$\pm$0.3\\
LBG-$\beta$4 &  9.70$\pm$2.23  & 112$\pm$25    &  30.9$\pm$8.0  & 3.2$\pm$0.6 & 1.3$\pm$0.4\\
LBG-$\beta$5 &  11.44$\pm$3.02 & 131$\pm$33    &  37.9$\pm$12.6 & 3.4$\pm$0.8 & 1.2$\pm$0.3\\

\hline
\multicolumn{6}{c}{\textbf{Stacking as a function of stellar mass (LBG-$M$)}}\\
\hline

LBG-M1 & 2.04$\pm$0.51 &  28.7$\pm$5.7     & 5.4$\pm$1.6  & 1.6$\pm$0.6 & 1.5$\pm$0.4\\
LBG-M2 & 4.95$\pm$1.05   &  61.2$\pm$11.6 & 11.6$\pm$2.9     & 2.2$\pm$0.6 & 1.5$\pm$0.2\\
LBG-M3 & 7.13$\pm$1.49  &  85.6$\pm$14.7  & 14.7$\pm$3.2     & 2.5$\pm$0.5 & 1.7$\pm$0.2\\
LBG-M4 & 10.1$\pm$2.1  &  117$\pm$22  & 21.2$\pm$4.6     & 2.8$\pm$0.5 & 1.7$\pm$0.2\\
LBG-M5 & 17.7$\pm$3.9  &  200$\pm$38  & 37.2$\pm$8.5    & 3.4$\pm$0.6 & 1.8$\pm$0.2\\
LBG-M6 & 31.6$\pm$6.3 &  349$\pm$73   & 72.6$\pm$16.5   & 4.1$\pm$0.6 & 1.8$\pm$0.2\\

\hline
\end{tabular}}

 \label{TabLFs}
\end{table}

%\bigskip
\subsubsection{SFR - stellar mass relation}

Many recent studies have found evidence that the SFR in galaxies correlates with stellar mass along a main sequence (MS) relation which evolves with redshift and represents a ‘steady’ mode of SF. A linear relation seems to well represent the main sequence at intermediate masses, however there are studies showing that the slope of this relation varies with the stellar mass or which different selections (e.g. \citealt{Guzman1997}; \citealt{Brinchmann00}; \citealt{Bauer2005}; \citealt{Bell05}; \citealt{Papovich06}; \citealt{Reddy2006}; \citealt{Daddi2007a}; \citealt{Elbaz2007}; \citealt{Noeske2007b}; \citealt{Pannella2009}; \citealt{Peng2010}; \citealt{Rodighiero2010a,Rodighiero2011}; \citealt{Karim2011}; \citealt{Whitaker2012, Whitaker2014}; \citealt{Rodighiero2015}; \citealt{Schreiber2015}; \citealt{Lee2015}; \citealt{Ilbert2015}).

Fig. \ref{sfr_mass} shows the average SFR - $M_{*}$ relation computed for our stacked LBGs as a function of $M_{*}$, along with previous stacking studies and observed relations (references on the Fig. \ref{sfr_mass}). Our results are in excellent agreement with the stacking at 1.4~GHz by \cite{Karim2011} and the stacking in far-IR by \cite{Schreiber2015} using a complete sample of star-forming galaxy. The far-IR stacking for UV selected galaxies by \cite{Heinis2014} presents higher SFRs than ours. It is interesting to note that we agree for the largest mass bin with the stacking at 850$\mu$m of LBGs by \cite{Coppin2015}, however their lower mass bins show higher SFR. We propose that the differences between our results and theirs come from the absence of correction for clustering in their stacking analysis. This matches with the fact that they over estimates the fluxes of their stacked sources, where the relative contribution of clustering get important in the faint sources. We would like to stress here that perform a careful stacking analysis is very important. The different between the results from \cite{Coppin2015} and ours is an illustration of the fact that a statistically-controlled stacking analysis is necessary to obtain reliable results.

The SFR-mass relation that we observe is well described by a power law with an average slope of 0.81$\pm$0.09 in our mass range. We are in excellent agreement with the last results ($0.8\pm0.08$) by \cite{Pannella2015} using a complete sample of star-forming galaxy.  We are also within the uncertainties with previous studies of individual LBGs detected in IRAC \citep{Magdis2010a} at $z\sim3$ and those lower redshift relations \citep{Daddi2007a, Elbaz2007, Peng2010} but using a higher normalization factor, where they found a slope of $\sim$0.9. We do not clearly detect any flattening of this relation at high $M_{*}$ \citep{Lee2015,Ilbert2015}. We are also in good agreement with \cite{Schreiber2015} who showed this dependence for all redshifts in their analysis.  

   \begin{figure}[h]
   \centering
   \includegraphics[width=\hsize]{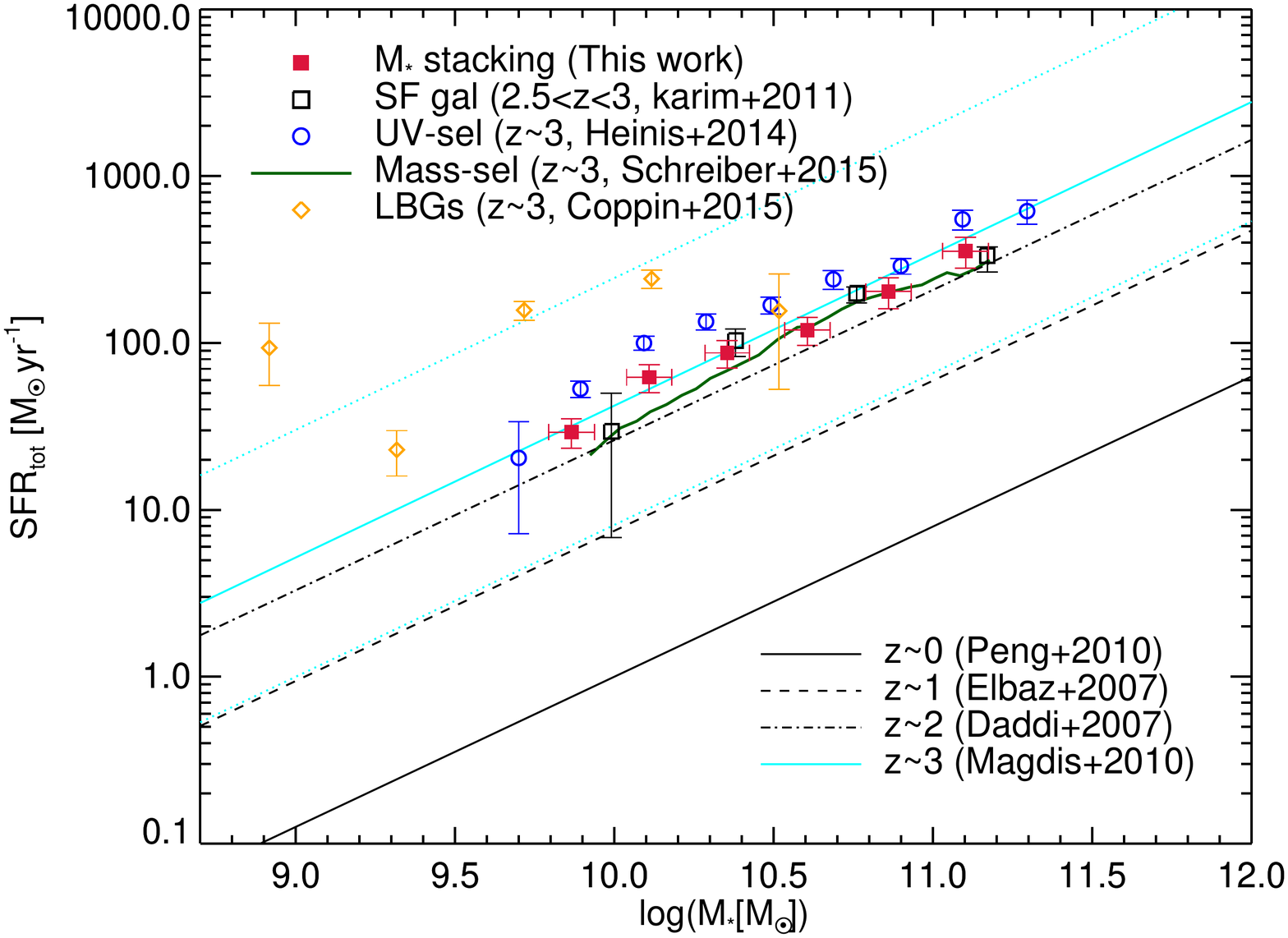}
      \caption{Total SFR versus stellar mass for our stacking sample as a function of stellar mass. The red filled squares show our results. We compare them with previous stacking studies using different sample selections at redshift $z\sim3$ for star-forming galaxies (\citealt{Karim2011} and \citealt{Schreiber2015}), UV selected galaxies \citep{Heinis2014} and LBGs \citep{Coppin2015}, all of them are plotted with open symbols. The various lines show previous observed sequences at $z=0$ \citep{Peng2010}, $z=1$ \cite{Elbaz2007}, $z=2$ \cite{Daddi2007a} and $z=3$ (from \citealt{Magdis2010a}, based on IRAC-detected LBGs). When necessary, the SFRs and stellar masses values have been converted to a \cite{Chabrier2003} IMF.}
         \label{sfr_mass}
   \end{figure} 
   
  \subsubsection{Difference between total SFR and corrected SFR by IRX-$\beta_{\mathrm{UV}}$ relation}\label{sfr_comp}

We investigate here the differences between the total SFR obtained from UV+IR by stacking a population of galaxies and with SFR corrected by the mean IRX-$\beta_{\mathrm{UV}}$ relation. We calculate for each individual object the corrected SFR by SFR$_{\mathrm corr}$ = SFR$_{\mathrm{UV}} \times 10^{0.4 A_{FUV}}$, where $A_{FUV} = 3.15+1.47~\beta_{\mathrm{UV}}$. Then, we split the sample in the same bins than the stacking as a function of $M_{*}$, and we compute the average of the SFR$_{\mathrm corr}$ values. Fig. \ref{sfr_mass_beta} shows the comparison between the mean of the two SFR estimators for each bin in stellar mass and the SFR$_{\mathrm corr}$ calculate for each LBGs. We obtain that the SFR$_{\mathrm corr}$ presents lower SFR values for $\log(M_{*}[M_{\sun}]) > 10$ than SFR$_{\mathrm total}$. The differences increase for higher stellar mass. However, we find a good agreement for both SFR estimator in the lowest stellar mass bins. This confirms the conclusion in Sect. \ref{IRX_beta_mass} that the SFR corrected form IRX-$\beta_{\mathrm{UV}}$ relation under estimates the SFR of the population of galaxies with high stellar mass, as \cite{Pannella2015}. But, it can be a good estimator of the average SFR for a populations of galaxies with low stellar mass. %Note that we already show that the IRX-$M_{*}$ relation present lower dispersion and can give more accurate correction of the dust attenuation for LBGs at $z\sim3$.

The IRX relations derived in Sect. \ref{att} are obtained from an average population of galaxies. As already shown, these relations present a dispersion due to the different nature of the galaxies. Therefore, their use could be useful but limited to  works with average population of galaxies, Applying them to individual galaxies can lead to strong under or over estimation of the true dust attenuation and therefore SFR. We also show that the selection of the sample might impact the resulting dust attenuation in the IRX planes, this means that depending on the sample selection ($M_{*}$, $\beta_{\mathrm{UV}}$, $L_{\mathrm{UV}}$) and on the science objectives the most valid dust attenuation correction must be chosen.% However, we demonstrate that the IRX-$M_{*}$ relation is less dispersed than the IRX-$\beta_{\mathrm{UV}}$ one. For our present sample of LBGs,  the best dust attenuation indicator for our sample is the $M_{*}$.

  \begin{figure}[h]
   \centering
   \includegraphics[width=\hsize]{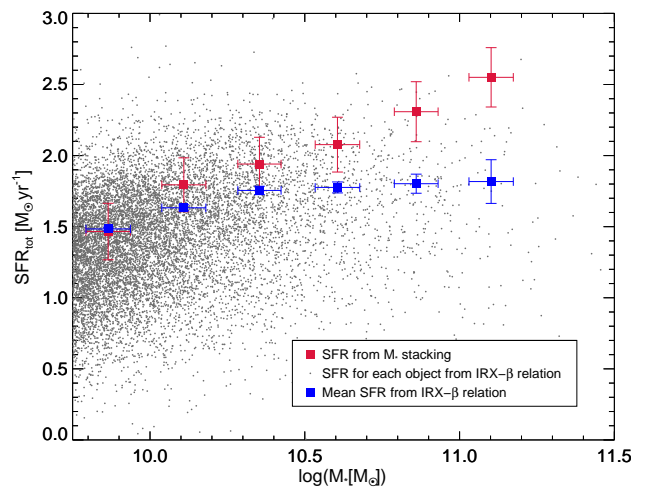}
      \caption{Comparison between SFR$_{total}$ and SFR$_{corr}$ as a function of the $M_{*}$. The red filled squares show the SFR$_{total}$ from the stacking results as a function of the $M_{*}$. The gray point are the individual SFR$_{corr}$ values for each LBG of our sample. The blue filled squares show the average of the SFR$_{corr}$, the error bars represent the error on the mean ($\sigma/\sqrt{N_{objects}}$).}
         \label{sfr_mass_beta}
   \end{figure} 

\section{SUMMARY AND CONCLUSIONS}

We presented in this paper a stacking analysis to study the rest-frame far-IR average properties of LBGs at $z\sim3$. We combine the available COSMOS multi-wavelength dataset to select a large sample of LBGs (around 22,~000) using the dropout technique and photo-$z$. Thanks to the large number of objects included into our sample, we are able to split it in several bins allowing us to explore the evolution of the dust attenuation and star formation rate as a function of the galaxy properties ($L_{\mathrm{FUV}}$, $\beta_{\mathrm{UV}}$ and $M_{*}$). We perform a stacking analysis in PACS (100 and 160~$\mu$m) images from the PACS Evolutionary Probe (PEP) team, SPIRE (250, 350 and 500~$\mu$m) images from the Herschel Multi-Tiered Extragalactic Survey (HerMES) programs and the AzTEC (1.1 mm) image from the ASTE, obtaining an average flux density for the population in each bin and band. Our main results can be summarized as follows

\begin{enumerate}
    \item We compute the full infrared SEDs and we derive the average IR luminosity for our LBGs as a function of their $L_{\mathrm{FUV}}$, $\beta_{\mathrm{UV}}$ and $M_{*}$. The obtained IR luminosities cover the range 3 $\times$ 10$^{10}$ to 3.23 $\times$ 10$^{12}$ $L_\odot$. We obtain that most of the stacked LBGs present an average $L_{\mathrm{IR}}$ similar to LIRGs, but the most massive and redder ones are ULIRGs. We find a power law correlation between the  average $L_{\mathrm{IR}}$ and the parameters $L_{\mathrm{FUV}}$, $\beta_{\mathrm{UV}}$ and $M_{*}$ that we use to split the sample in each stacking analysis.

    \item The average IRX (or dust attenuation) appears to be correlated with the $\beta_{\mathrm{UV}}$ slope. The relation follows the equation, $A_{FUV} = (3.15\pm0.12) + (1.47\pm0.14) ~\beta_{\mathrm{UV}}$. It lies below the relation derived from local starburst galaxies by M99, but is in agreement with the correction to the M99 relation by T12, suggesting that compact blue starbursts chosen by M99 provide a good estimate of the dust attenuation for our average LBG population.
    
    \item  The average IRX (or dust attenuation) is correlated with the $M_{*}$, following the relation: $\log \left( IRX \right) =(0.84\pm0.11) \log \left( M_{*}/10^{10.35}\right) + 1.17\pm0.05$. The IRX-$\beta_{\mathrm{UV}}$ and IRX-$M_{*}$ relations present an inconsistency between them, the more massive LBGs do not follow the IRX-$\beta_{\mathrm{UV}}$ relation and they show large dust attenuation and bluer colour.
    
	\item   We perform a stacking analysis where we split our LBG sample in the ($M_{*}$, $\beta_{\mathrm{UV}}$) plane in two different ways. The objective of these new stackings is to study the dispersion in the IRX-$\beta_{\mathrm{UV}}$ and IRX-$M_{*}$ planes and the inconsistency between the mean IRX-$\beta_{\mathrm{UV}}$ and IRX-$M_{*}$ relations. The IRX-$\beta_{\mathrm{UV}}$ plane presents a large dispersion due to differences in $M_{*}$. The dust attenuation shows a departure by up to 2.8 mag above the mean IRX-$\beta_{\mathrm{UV}}$ relation, for the same $\beta_{\mathrm{UV}}$ bin when the $\log(M_{*}[M_{\sun}])$ increases from 9.75 to 11.5. The IRX-$M_{*}$ plane is less dispersed, the dust attenuation also shows a departure by up to 1.3 mag above the mean IRX-$M_{*}$ relation, for the same $M_{*}$ bin when $\beta_{\mathrm{UV}}$ increases from -1.7  to 0.5. However, the low stellar mass LBGs ($\log(M_{*}[M_{\sun}]) < 10.5$) and red $\beta_{\mathrm{UV}}$ ($\beta_{\mathrm{UV}} > -0.7$), 15\% of the total sample, present larger dust attenuation than the mean IRX-$M_{*}$, but they are in agreement with the mean IRX-$\beta_{\mathrm{UV}}$ relation.

\item We suggest that we have to combine both, IRX-$\beta_{\mathrm{UV}}$ and IRX-$M_{*}$, relations to obtain the best estimation of the dust attenuation from the UV and NIR properties of the galaxies ($L_{\mathrm{FUV}}$, $\beta_{\mathrm{UV}}$, $M_{*}$).

   \item The  IRX (and dust attenuation) is roughly constant over the $L_{\mathrm{FUV}}$ range for the average population of LBGs, with a mean of 7.9 (1.8 mag). The IRX-$L_{\mathrm{FUV}}$ plane presents a large dispersion if we compare with the results in the stacking as a function of  $\beta_{\mathrm{UV}}$ and $M_{*}$ that can reach up to 2 magnitudes.
    
	\item The average SFR-$M_{*}$ relation is well approximated by a power law, with a slope of 0.81$\pm$0.09 in our stellar mass range. We show that our LBG sample is consistent with the main sequence of star formation. 
	
	\item If we compare the total SFR (IR+UV) obtained using stacking analysis and the one calculated by correcting the $L_{\mathrm{FUV}}$ using the IRX-$\beta_{\mathrm{UV}}$ relation (SFR$_{corr}$), we demonstrate that the IRX-$\beta_{\mathrm{UV}}$ relation underestimates the SFR for high stellar mass LBGs, but it gave a good estimation for lower stellar mass LBGs. 
	
	    \item The above relations provide phenomenological recipes to correct the observed $L_{\mathrm{FUV}}$ for dust attenuation, given a stellar mass or a $\beta_{\mathrm{UV}}$. These recipes are useful in the absence of observed far-infrared data. However, we stress that, even if they provide fairly good estimates of the amount of dust attenuation, the results must be seen as statistical and valid for average populations of galaxies. The methods should not be applied to individual galaxies without assuming large uncertainties in the result. 
	
\end{enumerate}

%These results will give us the possibility to model the SEDs of LBGs from UV to far-IR as a function of $\beta_{\mathrm{UV}}$ and $M_{*}$. We will be able to provide templates for star-forming galaxies at redshift $z\sim3$, which could be used to study and model higher redshift galaxies. 

\begin{acknowledgements}
PACS has been developed by a consortium of institutes led by MPE (Germany) and including UVIE (Austria); KU Leuven, CSL, IMEC (Belgium); CEA, LAM (France); MPIA (Germany); INAF-IFSI/ OAA/OAP/OAT, LENS, SISSA (Italy); IAC (Spain). This development has been supported by the funding agencies BMVIT (Austria), ESA-PRODEX (Belgium), CEA/CNES (France), DLR (Germany), ASI/INAF (Italy), and CICYT/MCYT (Spain). SPIRE has been developed by a consortium of institutes led by Cardiff Univ. (UK) and including: Univ. Lethbridge (Canada); NAOC (China); CEA, LAM (France); IFSI, Univ. Padua (Italy); IAC (Spain); Stockholm Observatory (Sweden); Imperial College London, RAL, UCL-MSSL, UKATC, Univ. Sussex (UK); and Caltech, JPL, NHSC, Univ. Colorado (USA). This development has been supported by national funding agencies: CSA (Canada); NAOC (China); CEA, CNES, CNRS (France); ASI (Italy); MCINN (Spain); SNSB (Sweden); STFC, UKSA (UK); and NASA (USA). The authors acknowledge financial contribution from the contracts PRIN-INAF 1.06.09.05 and ASI-INAF I00507/1 and I005110. The data presented in this paper will be released through the Herschel Database in Marseille (HeDaM; \url{http://hedam.oamp.fr/HerMES}). E. Ibar acknowledges funding from CONICYT/FONDECYT postdoctoral project N$^\circ$:3130504. We gratefully acknowledge the contributions of the entire COSMOS collaboration consisting of more than 100 scientists. This work makes use of TOPCAT (\url{http://www.star.bristol.ac.uk/?mbt/topcat/}).

\end{acknowledgements}

\bibliographystyle{aa}
\bibliography{Bibliography}

 \end{document}